\documentclass[12pt,preprint]{aastex}  

\def\ang{\AA}
\def\arcsec{\hbox{$^{\prime\prime}$}}

\def\gapprox{\lower.4ex\hbox{$\;\buildrel >\over{\scriptstyle\sim}\;$}}
\def\lapprox{\lower.4ex\hbox{$\;\buildrel <\over{\scriptstyle\sim}\;$}}

\def\ref#1{\par\noindent\hangindent1cm {#1}}

\shortauthors{ASCHWANDEN AND SCHRIJVER 2011}
\shorttitle{SOC}

\begin{document}

\title{		Coronal Loop Oscillations Observed with AIA - 
		Kink-Mode with Cross-Sectional and Density Oscillations}

\author{        Markus J. Aschwanden and Carolus J. Schrijver}

\affil{         Lockheed Martin Advanced Technology Center,
                Solar \& Astrophysics Laboratory,
                Org. ADBS, Bldg.252,
                3251 Hanover St.,
                Palo Alto, CA 94304, USA;
                e-mail: aschwanden@lmsal.com}

\author{	}

\affil{ 	}

\begin{abstract}
A detailed analysis of a coronal loop oscillation event is presented, 
using data from the Atmospheric Imaging Assembly (AIA) onboard the
Solar Dynamics Observatory (SDO) for the first time. The loop oscillation
event occurred on 2010 Oct 16, 19:05-19:35 UT, was triggered by an M2.9
GOES-class flare, located inside a highly inclined cone of a narrow-angle CME.
This oscillation event had a number of unusual features: (i) Excitation
of kink-mode oscillations in vertical polarization (in the loop plane);
(ii) Coupled cross-sectional and density oscillations with identical periods;
(iii) no detectable kink amplitude damping over the observed duration of
four kink-mode periods ($P=6.3$ min); (iv) multi-loop oscillations with
slightly ($\approx 10\%$) different periods; and (v) a relatively cool loop 
temperature of $T\approx 0.5$ MK. We employ a novel method of deriving
the electron density ratio external and internal to the oscillating loop
from the ratio of Alfv\'enic speeds deduced from the flare trigger delay
and the kink-mode period, i.e., $n_e/n_i=(v_A/v_{Ae})^2=0.08\pm0.01$. 
The coupling of the kink mode and cross-sectional oscillations can be explained
as a consequence of the loop length variation in the vertical polarization mode.
We determine the exact footpoint locations and loop length with
stereoscopic triangulation using STEREO/EUVI-A data. We model the
magnetic field in the oscillating loop using HMI/SDO magnetogram data
and a potential field model and find agreement with the seismological
value of the magnetic field, $B_{kink}=4.0\pm0.7$ G, within a factor
of two. 
\end{abstract}

\keywords{Sun: Flares --- Sun : Corona --- Sun: Extreme Ultra-Violet (EUV)
                      --- Sun : oscillations --- Sun : waves      }

\section{       INTRODUCTION 			}

Propagating waves and standing waves (eigen-modes) in coronal plasma
structures became an important tool to probe the physical parameters, 
the dynamics, and the magnetic field in the corona, in flare sites,
and in coronal mass ejections (CMEs). Recent reviews on the theory and
observations of coronal seismology can be found in Roberts and Nakariakov 
(2003), Erdelyi et al.~(2003), Roberts (2004), Aschwanden (2004, 2006), 
Wang (2004), Nakariakov and Verwichte (2005), Banerjee et al.~(2007),
Andries et al.~(2009), Ruderman and Erdelyi (2009), and Taroyan and Erdelyi 
(2009). Substantial progress was accomplished
in applying MHD wave theory to coronal observations with previous 
instruments, such as the discovery of global waves with EIT/SOHO 
(Thompson et al.~1998, 1999), fast kink-mode loop oscillations with TRACE 
(Aschwanden et al.~1999; Nakariakov et al.~1999), fast sausage mode
oscillations in radio wavelengths (Roberts et al.~1984; Asai et al.~2001;
Melnikov et al.~2002; Aschwanden et al.~2004), 
slow (acoustic) mode oscillations with SUMER/SOHO
(Wang et al.~2002; Kliem et al.~2002), slow (acoustic) propagating waves
with UVCS/SOHO (Ofman et al.~1997) and Hinode (Erdelyi and Taroyan 2008), 
EIT/SOHO (DeForest and Gurman 1998),
and TRACE (De Moortel et al.~2002a,b), fast Alfv\'enic waves with SECIS
(Williams et al.~2001; Katsiyannis et al.~2003), or fast kink waves
with TRACE (Verwichte et al.~2004; Tomczyk et al.~2007). 
Sausage oscillations observed and identified directly in cross-sectional
area change of a solar magnetic flux tube are reported by Morton et al.~(2001).
However, temporal cadence of space-borne EUV imagers (such as EIT/SOHO, TRACE,
STEREO) was mostly in the order of 1-2 minutes, which is just at the
limit to resolve fast MHD mode oscillations (with typical periods of
3-5 minutes) and is definitely too slow to resolve or even detect
fast MHD waves that propagate with Alfv\'enic speed. An Alfv\'en wave
with a typical coronal speed of $v_A \approx 1000$ km s$^{-1}$ traverses
an active region in about 1 minute. With the advent of the Atmospheric
Imaging Assembly (AIA) onboard the Solar Dynamics Observatory (SDO), which 
provides a permanent cadence of 12 s, we have an unprecedented opportunity 
to study the exact timing of the excitation mechanisms of coronal MHD
waves and oscillations, which often involve an initial impulsive pressure 
perturbation in a flare and CME source site, that launches various MHD
waves and oscillations in surrounding resonant coronal structures 
(loops, fans, CME cones, and cavities). 

Here we conduct a first AIA study on a
coronal loop oscillation event, observed on 2010 October 16, which
exhibits a favorable geometry, unobstructed view, prominent undamped
oscillations, an unusual coupling of kink-mode and cross-sectional
(and density) oscillations (not noticed earlier), 
and a rare case of vertical kink-mode polarization. 
In Section 2 we present various aspects of the
data analysis and modeling, while theoretical and interpretational
aspects are discussed in Section 3, with the major findings and conclusions
summarized in Section 4. 

\section{       DATA ANALYSIS 			}

\subsection{	Instrument 			}

The {\sl Atmospheric Imaging Assembly (AIA)} instrument onboard the {\sl 
Solar Dynamics Observatory (SDO)}
started observations on 2010 March 29 and produced since then 
continuous data of the full Sun with four $4096 \times 4096$ detectors
with a pixel size of $0.6\arcsec$, corresponding to an effective 
spatial resolution of $\approx 1.6\arcsec$. AIA contains 10 different 
wavelength channels, three in white light and UV, and seven EUV channels, 
whereof six wavelengths (131, 171, 193, 211, 335, 94 \ang )  
are centered on strong iron lines (Fe VIII, IX, XII, XIV, XVI, XVIII),
covering the coronal range from $T\approx 0.6$ MK to $\gapprox 16$ MK.
AIA records a full set of near-simultaneous images in each temperature
filter with a fixed cadence of 12 s. Instrumental descriptions 
can be found in Lemen et al.~(2011) and Boerner et al.~(2011).

\subsection{	Observations and Location 			}

A major flare of GOES class M2.9 occurred on 2010 Oct 16, 19:07-19:12 UT
at location W26/S20 ($+390\arcsec$ west and $-410\arcsec$ south 
of Sun center),
which triggered a number of loop oscillations in the westward direction of
the active region (NOAA 1112). In this study we focus on the detailed
analysis of a loop at the loop apex position $+698\arcsec$ west and
$-243\arcsec$ south of Sun center, which displays prominent oscillations.
The location of this loop with respect to the flare center, is shown in Fig.~1.
The oscillating loop is discernible as a faint semi-circular structure
in the logarithmically-scaled intensity image in 171 \ang\ (Fig.~1, top
panel), or even clearer in the difference image (19:22:36 UT - 19:21:00 UT)
in Fig.~1 (bottom panel), where the times were chosen at the maximum and
subsequent minimum of an oscillation period. The 171 \ang\ intensity
image (Fig.~1) shows also irregular moss-like structure in the background of 
the oscillating loop of similar brightness, which poses some challenge for 
exact measurements of the loop oscillation parameters, because the
background is time-variable, even on the time scale of the oscillation period. 

\subsection{	Transverse Loop Oscillations			}

Loop oscillations are traditionally investigated most easily in
time-difference movies. [Movies of this oscillation event in 171 \ang\
intensity and running-difference format are available as supplementary
data in the electronic version of this journal]. However, a variety
of time-differencing schemes can be applied in order to enhance the best
contrast. We explore a variety of time-differencing schemes in Fig.~2,
for a data stripe oriented perpendicular to the loop axis at its apex
with a length of 30 pixels and a width of 10 pixels (indicated with
a small rectangle in Fig.~1). 
We construct time-slice plots with $n_t=150$ time frames on the x-axis
(covering the time interval from 19:05 UT to 19:35 UT with a
cadence of $\Delta t = 12$ s) and a spatial dimension in direction 
transverse to the loop on the y-axis (with $n_y=30$ pixels), averaged
over the $n_w=10$ pixels of the stripe width (parallel to the loop). 
We show 5 different differencing
schemes of this time slice in Fig.~2, using a high-pass filter
(Fig.~2 top panel), a baseline difference (Fig.~2, second panel),
and a one-sided (Fig.~2, third panel), a symmetric (Fig.~2, fourth panel),
and a running-minimum difference scheme (Fig.~2, bottom panel), 
which is defined as
\begin{equation}
	\Delta F(t_i,y_j)=F(t_i,y_j)-min[F(t_{i-k},y_j),...,F(t_{i+k},y_j)] 
	, \
\end{equation}
so it subtracts a running minimum evaluated within a time interval with
a length of $2k$ pixels symmetrically placed around every time slice.
Each method has its merits and disadvantages, as it can be seen in Fig.~2.
The biggest challenge is the non-uniformity and time variability of the 
background. An 
additional complication is the presence of fainter secondary oscillating 
loops, which appear like ``echoes'' in the time-slice plots. For further
analysis we adopt the running-minimum differencing scheme (Fig.~2, bottom), 
which appears to have the best signal-to-noise ratio of the oscillating 
features.

The measurement of the loop oscillation amplitude variation $a(t)$ as a function
of time $t$ can be done (i) by localizing the cross-sectional flux maxima
in running-difference time-slice plots, (ii) by cross-correlation of subsequent
time slices, or (iii) by fitting a Gaussian profile to the cross-sectional
flux profiles. We find that the first and the latter method are most
robust. From the running-minimum time-slice plot (Fig.~3 top frame)
we perform fits of Gaussian profiles $F_{fit}(s,t)$ to the observed
cross-sectional flux profiles $F(s,t)$ in each time slice $t$ (using
the standard GAUSSFIT.PRO routine in the IDL software),
\begin{equation}
	F_{fit}(s,t)=f(t) \exp{ \left( -{[s - a(t)]^2 
	\over 2 \sigma_s^2(t)} \right)} +b(t)
	\ ,
\end{equation}
which yields the four coefficients of the peak flux $f(t)$, the oscillation
amplitude $a(t)$, Gaussian width $\sigma_s(t)$, and mean background flux 
$b(t)$ for each time $t$. The 4-parameter fits 
The cross-sectional flux profiles $F(s,t)$ and the Gaussian fits $F_{fit}(s,t)$
are shown in Fig.~4 for each time in the interval between
$t_{1}$=19:05 UT and $t_{150}$=19:35 UT, while a corresponding
time-slice with Gaussian fits is rendered in color scale in Fig.~3
(second panel). The average Gaussian loop width during the oscillation
period is $\sigma_s=2.1$ Mm, which corresponds to a FWHM loop width of
$w=\sigma_s 2 \sqrt{2 \ln 2}=4.9\pm0.6$ Mm. 

We are fitting now a sinusoidal function with a linear drift
to the location of the oscillation amplitudes $a(t)$ (crosses in Fig.~3, 
fourth panel), using the Powell optimization routine (Press et al.~1986)
from the IDL software,  
\begin{equation}
	a_{fit}(t)=a_0 + a_1 \sin{\left( {2 \pi (t-t_0) \over P} \right)}
		  +a_2 {(t-t_0) \over P} \ ,
\end{equation}
for which we find a midpoint position $a_0=6.3$ Mm, a drift velocity
$a_2/P=0.8$ km/s, an oscillation period of $P=395$ s (6.4 min),
an oscillation amplitude $a_1=1.8$ Mm, and a sinusoidal onset time of
$t_0=393$ s after the start of the time slice at 19:05:00 UT, i.e.,
at 19:11:33 UT. The onset time of the oscillation will be important 
to measure the exciter speed of the trigger.
The fit of the sinusoidal amplitude function $a_{fit}(t)$ to the measured
amplitude $a(t)$ is shown in Fig.~3 (fourth panel). The fitted function
with a constant amplitude $a_1$ appears to be appropriate for the
duration of $N_{pulse}=(t_{150}-t_{33})/P=1407/395=3.6$ oscillation 
periods, since we do not observe any significant damping of the amplitude 
during this time interval.  

\subsection{	3D Loop Geometry 		}

The projected loop shape is close to a semi-circular geometry
(Fig.~1, bottom), and thus we can assume that the loop plane
is near the plane-of-sky or nearly perpendicular to the line-of-sight.
The location of the loop curvature center is at a distance of
$\approx 740\arcsec$ from Sun center or 0.77 solar radii, which corresponds
to a heliographic angle of $\alpha=50^\circ$ from disk center. 

The full 3D geometry of the loop can be obtained from the combination
of the EUVI instrument onboard STEREO and AIA observations, a procedure 
that we carry out for the first time here. 
The loop was in the field-of-view of STEREO/A(head) at this
time, but was occulted for STEREO/B. The STEREO/A spacecraft was located
on 2010 Oct 16 at a separation angle of $\alpha_A=83.583^\circ$ to the 
east of Earth, at a latitude of $\beta_A=-0.119^\circ$ from the Earth 
ecliptic plane. In Fig.~5 we show nearly contemporaneous AIA and
EUVI/A difference images of the loop, which clearly show the
oscillatory motion of the loop, after highpass-filtering of the EUVI/A
image. Unfortunately, EUVI/A observed only in a different wavelength 
of 195 \ang\ at this time, while the oscillation is best visible in the
171 \ang\ channel in AIA. EUVI/A had also a lower cadence ($\approx 5$ min
vs. 12 s in AIA) and the spatial resolution
of EUVI ($1.6\arcsec$ pixels) is about three times coarser than AIA
($0.6\arcsec$ pixels). Nevertheless, the image quality is sufficient to
approximately determine the 3D loop geometry. We subtracted the earlier 
image from the later image, and thus a density increase in the difference
images (white in Fig.~5) indicates an inward loop motion (in the
AIA image) and a correlated density compression (in the EUVI/A image). 
We rotate the 2D-coordinates of the loop traced in AIA (Fig.~5 left) 
into the coordinate system of EUVI/A with variable heights and inclination
angle of the loop plane. By matching the position and direction of the
loop ridge in EUVI/A we obtain the absolute height range of the traced
loop segment, i.e., $21.7 < h_{segm} < 37.4$ Mm. In order to locate the
positions of the footpoints we extrapolate the traced loop segment in
both directions and define the positions of the loop footpoints where
the coplanar extrapolation intersects with a height $h=0$ above the solar
surface. The so-defined extrapolated footpoint positions are found at
$F_1=(685\arcsec, -305\arcsec)$ (south of traced loop) and 
$F_2=(615\arcsec, -268\arcsec)$ (east of trace loop) 
with respect to Sun center (Fig.~5). The apex or midpoint of the traced
loop segment (at $s=L_{loop}/2$) is located at position 
$(x_{apex},y_{apex})=(698\arcsec, -243\arcsec)$, for which we show
time-slice plots of the oscillation in Figs.~2-4 (i.e., segment \#6 
in Fig.~6). The apex location will also be used to define the arrival
time of the exciting wave and starting time of the loop oscillation
in Section 2.11. The inclination angle of the loop plane
to the local vertical is found to be 
$\vartheta \approx 20^\circ \pm 20^\circ$, but cannot be determined
more accurately because of the short loop segment detectable in EUVI/A. 

From the absolute 3D coordinates $(x_i,y_i,z_i), i=1,...,n$ of the 
stereoscopically triangulated loop we can calculate the full loop 
length $L_{loop}$,
\begin{equation}
	L_{loop} = \sum_{i=0}^{n-1} \sqrt{ \left[(x_{i+1}-x_i)^2+
		(y_{i+1}-y_i)^2+
		(z_{i+1}-z_i)^2 \right]} \ ,
\end{equation}
for which we find $L_{loop}=163$ Mm. The traced loop segment over
which amplitude oscillations are clearly visible covers the fraction
from $s_1/L=0.23$ to $s_2/L=0.78$ of the total loop length and has
only a length of $L_{segm}=123$ Mm. If we approximate the 3D loop geometry
with a coplanar semi-circular shape, the loop curvature radius is estimated
to be $r_{loop} \approx L_{loop}/\pi = 52$ Mm. 

The plane of transverse loop oscillations with respect to the average
loop plane cannot accurately be determined with the existing STEREO data,
but they are roughly coplanar, based on the centroid motion constrained
by AIA that is absent in EUVI/A from a near-perpendicular view (Fig.~5).
Coplanar kink mode oscillations corresponds to a vertical polarization.

\subsection{	Spatial Variation of Loop Oscillation		}

In a next step we analyze the spatial variation of the transverse
kink-mode oscillation $a(t)$ as a function of the spatial loop position,
which we specify with a segment number running from 
segment \#1 at the loop length coordinate $s_1/L=0.23$ (near the first
footpoint $F_1$) to segment \#10 at $s_2/L=0.78$ (near the second loop 
footpoint). This analysis serves a two-fold purpose: 
(1) to detect possible asymmetries of the kink mode, and (2) to detect
possible propagating waves.

In Fig.~6 we show the analysis of the loop oscillation of 10 different loop 
segments, numbered consecutively (\# 1-10) from the southern loop footpoint
$F_1$ to the north-eastern footpoint $F_2$ along the loop axis with
loop length coordinate $s$. The location and orientation of the 
time-slice stripes is indicated in the left panels of Fig.~6, the
running-minimum difference time slices are shown in the middle panels 
of Fig.~6, and the sinusoidal fits $a_{fit}(t)$ to the loop amplitudes in
the right-hand panels of Fig.~6, which also contains the best-fit parameters.
If we discard the the two noisiest
segments near footpoint $F_1$ (Segment \#1 and \#2 in Fig.~6), we obtain for
the others a mean amplitude of $\langle a_1(s) \rangle=1.8\pm0.4$ Mm, 
a mean period of $\langle P(s) \rangle =373\pm30$ s (6.2$\pm$0.5 min), 
and a mean starting time of $\langle t_0(s) \rangle =399 \pm 35$ s. 
Thus the variation of best-fit periods
and starting times is only $\approx 8\%$, and thus we 
conclude that there is no significant phase shift of the oscillation 
amplitude along the loop that could be considered as a propagating wave. 
Thus, we deal with a pure standing wave of the fast MHD kink-mode.

In Fig.~7 we show the spatial variation of the oscillation amplitude
$a_1(s)$ in the context of an intensity image (Fig.~7, top left) and
a running-minimum difference image (Fig.~7, top right). 
The locations of the 10 azimuthal time-slice stripes are shown
in Fig.~7 (bottom left), over which the amplitude oscillation were 
measured in Fig.~6. The dependence of the oscillation amplitude $a_1(s)$ 
along the loop shows a maximum amplitude of $a_1^{max} = 2.0$ Mm near the 
loop apex. A sinusoidal displacement along the loop axis is ideally
expected for a kink eigen-mode with fixed nodes (compare with the 
analogy of a violin string). Our measurements, however, rather show
a slightly distorted and asymmetric function, which can be 
approximated by a squared sine function (to account for the curvature
of the loop) and a nonlinear dependence $a(s^{0.75})$ on the loop length 
(to account for the asymmetry of the loop, as evident from the
stereoscopic triangulation of the footpoints, see Fig.~7 bottom right panel), 
\begin{equation}
	a_1(s) = a_1^{max} \sin^2{ \left[ \pi \left( {s \over L}
	\right)^{0.75} \right]} \ ,
\end{equation}
where $s=0$ and $s=L$ mark the nodes at the true footpoints $F_1$ and $F_2$. 
The observed oscillation amplitudes $a(s)$ follow the squared sine function
closely in the range of $s/L \lapprox 0.6$, but deviate in the range of 
$0.6 \lapprox s/L \lapprox 0.75$, probably because of the interference of 
a secondary oscillating loop. 

\subsection{	Multiple-Loop Oscillations 			}

From watching the AIA 171 \ang\ movies (see electronic supplementary data
to this paper) and from the time-slice plots shown in Figs. 2, 3, and 6
it appears that multiple loops are involved in kink-mode oscillations.
The previous analysis has determined the average dynamic parameters of
the collective ensemble of individual loop strands. In Figure 8 we are
fitting a 2-loop model to the time-slice plots obtained near the loop
apex, which yields slightly
different periods ($P_1=375$ s and $P_2=336$ s), oscillation amplitudes 
($a_1= 2.3$ Mm and $a_2=2.0$ Mm), and loop centroid positions 
($x_1=6.8$ Mm and $x_2=6.1$ Mm), but a common start time $t_0=423$ s
(i.e., 19:12:03 UT). Thus the loops are excited in phase, but the
secondary loop has an oscillation period that is about 10\% shorter.
The secondary loop seems also to have a shorter lifetime and is only
visible in the 171 \ang\ filter for about 2 oscillation periods (compared
with 3.6 periods of the primary loop.)

\subsection{	Intensity Modulation During Loop Oscillations	}

The 171 \ang\ intensity of the background-subtracted loop intensity exhibits
strong modulations, being strongest near the beginning, but fading out 
gradually at the end of the time interval of oscillations. We show in Fig.~3 
(bottom panel) the background-subtracted intensity flux profile $f(t)$ 
as measured near the loop apex from the Gaussian cross-sectional profile
fits (Eq.~2). Amazingly, the intensity flux modulation
appears to be in synchronization with the oscillation amplitude, which
is a very interesting property that we have not noticed in previous
observations of loop oscillations (e.g., in the 26 cases observed with
TRACE; Aschwanden et al.~1999, 2002). In fact, the loop flux
modulation $f(t)$ (Fig.~3, bottom panel) occurs in anti-phase to 
the amplitude modulation $a(t)$ that is measured in upward direction 
away from the loop curvature center. 
In addition to the oscillation-modulated
variation, the flux decays as a function of time, which can be described
with a linear decay rate $df/dt$, similar as found for 8 loops with kink-mode
oscillations during the 2001 Apr 15, 21:58 UT, flare (Aschwanden and
Terradas 2008). Thus, we fit a sinusoidal function with a linear
decay rate $f_2/P_f$,
\begin{equation}
	f_{fit}(t)
	= f_0 + f_1 \sin{\left( {2 \pi (t-t_f) \over P_f} \right)}
	      + f_2 {(t-t_f) \over P_f} \ .
\end{equation}
We find a peak flux of $f_0=61$ DN s$^{-1}$, a flux modulation of
$f_1 = 7.5$ DN/s, a linear decay rate of $f_2/P_f=-0.037$ (DN s$^{-2}$), 
which defines a loop lifetime of  
$t_{life}= P_f (f_0/f_2) = 1649$ s (27 min) and is compatible
to the loop cooling times $\tau_{cool}=17 \pm 7$ min found in
Aschwanden and Terradas (2008), modeled also in Morton and Erdelyi (2009,
2010). It is therefore suggestive to interpret
the observed lifetime of the oscillating loop as the detection time of
a loop that cools through the AIA 171 \ang\ passband. For the flux
modulation that is anti-correlated with the amplitude oscillation 
we suggest an interpretation in terms of density compression by
cross-sectional loop width oscillations, similar to a sausage mode,
which is modeled in the next section.

\subsection{	Density Modulation During Loop Oscillations 	}

In the previous Sections we established that the vertical oscillation 
amplitude amounts to $a_1^{max}/r_{loop}=2/52=4\%$. If we assume
that the loop is embedded in a magnetic field with a constant pressure
across the loop cross-section (to first order) in a low plasma
$\beta$-parameter environment, the ambient magnetic field lines are
expected to oscillate in synchronization with a displacement that is
proportional to the loop amplitude. A consequence of this scenario
is that the loop cross-sectional radius $r(t)-r0$ varies 
proportionally to the loop amplitude $a(t)$,
\begin{equation}
	r(t) = r_0 \left(1 + {a(t) \over r_{loop}}\right) \ ,
\end{equation}
leading to a modulation of the loop cross-sectional area 
$A(t)=\pi r^2(t)$ that scales quadratically to the loop radius 
$r(t)$, \begin{equation}
	A(t) = A_0 \left(1 + {a(t) \over r_{loop}}\right)^2 \ .
\end{equation}
Since the loop footpoints are anchored at fixed positions in the
photosphere, we can characterize the oscillating loop shape with
an elliptical geometry to first order, which oscillates around the
semi-circular geometry of the loop at rest, as depicted in Fig.~7.
The loop length of a half ellipse is mathematically (to first order),
\begin{equation}
	L = \pi {r_{minor} + r_{major} \over 2} + ... \ ,
\end{equation}
where $r_{minor}$ is the minor semi-axis and
$r_{major}$ is the major semi-axis of the ellipse. For the semi-circular
limit the radii are equal, $r_{minor}=r_{major}=r_{loop}$, and the half
loop length is $L= \pi r_{loop}$. Assigning the minor axis to the half
footpoint separation, $r_{minor}=r_{loop}$, and the major axis to the
vertical radius with a small oscillation amplitude, 
$r_{major}=r_{loop}+a(t)$, the elliptical loop length varies 
(to first order) as,
\begin{equation}
	L(t) = \pi \left( r_{loop} + {a(t) \over 2} + ... \right) 
	     = \pi r_{loop}  \left(1 + {a(t) \over 2 r_{loop}} + ... \right) \ .
\end{equation}
The volume of the loop, $V(t)$, varies then consequently with
the 5/2-power of the amplitude variation (to first order),
\begin{equation}
	V(t) = A(t) L(t) 
	     = V_0  \left(1 + {a(t) \over r_{loop}} + ... \right)^{5/2} \ .
\end{equation}
The electron density inside the loop, assuming particle conservation in
adiabatic compression and expansion processes, varies then reciprocally 
to the loop volume, 
\begin{equation}
	n_e(t) \propto V(t)^{-1} 
	     = n_0  \left(1 + {a(t) \over r_{loop}} + ... \right)^{-5/2} \ .
\end{equation}
For optically thin emission, as it is the case in EUV and soft X-rays for
coronal conditions, the flux intensity scales with the square of the
electron density times the column depth $dz$ (which is here assumed to be
proportional to the loop diameter $dz \propto \sqrt{A(t)}$), 
yielding an anti-correlation of the flux with the
4-th power of the amplitude oscillation, 
\begin{equation}
	F(t) \propto n_e^2(t) \ dz(t)
	     = F_0  \left(1 + {a(t) \over r_{loop}} + ... \right)^{-4} \ .
\end{equation}
Thus, the small-amplitude variation of $a_1^{max}/r_{loop}=2.2/52=0.042$
is amplified with the 4-th power,
\begin{equation}
	F_{max} = F_0  \left(1 - 4 {a_1^{max} \over r_{loop}} + ... \right) 
	\approx 1.18^{-1} \ ,
\end{equation}
which yields a flux modulation of 18\% with respect to the mean value $f_0$.
In Fig.~3 (bottom panel) we fitted the flux variation and found indeed
an average mean modulation factor of $q_f = f_1/\langle f_0(t) \rangle
 \approx 7.45/30) \approx 24\%$, for the average of the total flux 
$\langle f_0(t) \rangle \approx (61+10)/2 \approx 30$ DN s$^{-1}$ 
during the oscillatory episode. 
Thus our model predicts the correct time phase and approximate amount 
of oscillatory intensity flux modulation, which is anti-correlated to 
the sinusoidal loop amplitude oscillation (Fig.~9).
The MHD wave mode that is associated with cross-sectional variation is 
called sausage mode or symmetric $m=0$ mode of fast MHD waves
(e.g., Roberts 1984), which has a distinctly different 
eigen-mode period than the kink mode. The cross-sectional and density 
variation that were found in synchronization with the kink mode here 
(which has the same geometric and density properties as the sausage mode,
but a different period than predicted by the MHD dispersion relation), 
is a novel result of this study. 
This characteristic seems to be a particular property of oscillations 
in the loop plane (Fig.~7), 
also called ``vertical polarization of kink mode'' (Wang and Solanki 2004;
Verwichte et al.~2006a,b), which would not occur (to first order) for 
transverse oscillations in perpendicular direction to the loop plane.

\subsection{	Density and Temperature Analysis of Oscillating Loop	}

Having the 6 coronal AIA filters available that cover a temperature
range of $T \approx 0.6-16$ MK for the entire oscillation episode with
the same cadence of 12 s we are in an unprecedented position to conduct
an accurate diagnostics of the electron temperature and density of the
oscillating loop. For this purpose we extract loop-aligned subimages
in all 6 coronal wavelengths in 10 spatial segments ($s_i, i=1,...,10$)
at the loop locations as indicated in Fig.~7 (bottom left) and at
10 consecutive times ($t_j, j=1,...,10$) during the time interval of
19:05-19:35 UT. We show the $6 \times 10$ subimages for the 10 different
times for loop segment \# 6 near the loop apex in Fig.~10 (left half),
as well as the averaged cross-sectional loop profiles resulting from these
subimages in Fig.~10 (right half). We calculate also the cross-correlation
coefficients of these subimages with the simultaneous subimage in the
detected wavelength of 171 \ang\ (indicated by the numbers in each
subpanel in Fig.~10). From this information shown in Fig.~10 it is 
very clear that the oscillating loop exhibits a near-Gaussian
cross-sectional profile only in the 171 \ang\ filter, while the
131 and 193 \ang\ filters show only a mild correlation ($CCC \approx 
0.4-0.5$) and the remaining filters (211, 335, and 94 \ang ) are
absolutely uncorrelated ($CCC \approx 0.0-0.1$), which already narrows
down the loop temperature to peak response temperature of the 171 \ang\
filter at $T \approx 10^{5.9} \approx 0.8$ MK. 

In Fig.~11 (left side) we show the AIA temperature response functions,
where the low-temperature response of the 94 A filter is corrected by
an empirical factor of $q_{94}=6.7\pm 1.7$ (Aschwanden and Boerner 2011).
The total fluxes $f^{tot}(t)$ (histograms with error bars in Fig.~11
middle panels) and background fluxes $f^{back}(t)$ (hatched areas
in Fig.~11 middle panels) are also shown, where the background is
evaluated based on the Gaussian cross-sectional fits (Fig.~10).
The difference $f^{loop}(t)=f^{tot}(t)-f^{back}(t)$ is attributed
to the EUV flux of the oscillating loops and is modeled with a
single-Gaussian differential emission measure (DEM) distribution
by forward-fitting (according to the method described in Aschwanden
and Boerner 2011), 
\begin{equation}
	EM(T) = EM_0 \ \exp{\left( - 
	{(T-T_0)^2 \over 2 \sigma_T^2}\right)} \ ,
\end{equation}
with the best-fit DEM solutions shown in Fig.~11
(top right panel) for the 10 consecutive time steps. The single-Gaussian 
DEM fits yield an average peak temperature of $T_0=0.57\pm0.14$ MK and
a Gaussian temperature width of $\sigma_{log(T)}=0.18\pm0.10$
(Fig.~11, right side), which 
corresponds to a near-isothermal temperature distribution at the limit of the
temperature resolution $\Delta log(T) \approx 0.3$ of the AIA filters,
similarly as found for a statistical set of other loops analyzed from 
TRACE (Aschwanden and Nightingale 2005) or AIA (Aschwanden and Boerner 2011). 
The goodness-of-fit of the best-fit DEM solutions is found to be
$\chi^2_{red}=1.10\pm0.05$. The average agreement of the observed and
modeled fluxes is found to be $\lapprox 3\%$ in the three filters with
the highest fluxes (Fig.~11,
middle column). The largest relative deviation occurs in the 94 \ang\ filter,
which are known to have an incomplete temperature response function due to
missing lines of Fe X transition (Aschwanden and Boerner 2011).
Also we have to keep in mind that the largest flux deviations in the fits
are only in the order of $\approx 0.2-0.5$ DN/s in the 94, 131, and 335 \ang\
channels, which results mostly from uncertainties in the 
(time-variable) background evaluation rather than from the 
statistical photon noise.

Assuming a filling factor of unity, we can estimate the mean electron
density in the oscillating loop,
\begin{equation}
	n_0 = \sqrt{ EM_0 \over w} \ ,
\end{equation}
for which we obtain a mean value of $n_0 = (1.9\pm0.3) \times 10^8$
cm$^{-3}$, based on average loop widths of $w=4.9\pm0.6$ Mm
(Fig.~11, right side), which is measured near the apex of the loop 
for the segment \# 6 shown as cross-sectional loop profiles in Fig.~10. 

\subsection{	Radiative Cooling Time Scale		}

Since the issue has been raised that the lifetime of oscillating loops
(defined by the detection time in a given temperature filter) is
commensurable with the duration of an observed oscillation event
(Aschwanden and Terradas 2008; Morton and Erdelyi 2009),
let us explore whether the theoretically predicted time scales are
consistent with the observed flux decay. At the relatively low
coronal temperatures of $T_e \lapprox 1.0$ MK observed in EUV, radiative 
cooling is the dominant time scale, while conductive cooling is only
relevant at much hotter plasma temperatures in soft X-rays. 
Assuming an impulsive heating episode with subsequent cooling 
we can approximate the temperature evolution
with an exponentially decaying function over some temperature range,
\begin{equation}
        T_e(t-t_1) = T_e(t_1)\ \ 
	\exp{ \left[ - {(t-t_1) \over \tau_{cool}} \right]} \ ,
\end{equation}
where the temperature cooling time $\tau_{cool}$ corresponds to
the radiative cooling time $\tau_{rad}$, 
\begin{equation}
        \tau_{rad}(n_{0},T_{0})={9 \over 5} {k_B T_{0}^{5/3} \over n_{0} \Lambda_0}
	\ ,
\end{equation}
with $\Lambda_0 \approx 10^{-17.73}$ erg cm$^3$ s$^{-1}$
being the radiative loss rate at EUV temperatures ($T \approx 1.0$ MK).
For our measured values of $T_{0}=0.57$ MK and $n_{0}=1.9 \times 10^8$
cm$^{-3}$ at the loop apex we estimate $\tau_{rad} \approx 2750$ s
(46 min). The loop lifetime $\tau_{171}$ in the 171 \ang\ filter, 
which has a FWHM temperature range of $T_{171}=0.53-1.17$ MK,
is then $\tau_{171}=2200$ s (37 min) according to Eq.~17, 
which if fully consistent with the observed flux decay time
$\tau_{life}=P_f(f_0/f_2)=1650$ s (27 min) based on the fitted time
profile (Eq.~6) to the measurement shown in Fig.~3
(bottom panel). Therefore, we can interpret the observed flux decay
seen in the 171 \ang\ as a consequence of the radiative cooling time.
Based on this cooling scenario we would predict an initial temperature
of $T_e(t=t_1) \approx 0.8$ MK at the beginning of the oscillation event 
and a temperature of $T_e(t=t_2) \approx 0.4$ MK at the end of the
oscillation episode. The predicted temperature drop $T_e(t)$ for a
single loop cannot be retrieved by the DEM modeling (Fig.~11, left second 
panel), because the two nearly cospatial oscillating loops cannot be 
separated and thus we can only measure the combined emission 
measure-weighted temperature evolution of near-cospatial loops. 
Considering the analysis of the temperature evolution in other channels, 
only the 131 \ang\ channel has a temperature response to slightly lower 
values than the 171 \ang\ channel, but is about two orders of magnitude 
less sensitive and thus is unsuitable for a quantitative analysis. 

\subsection{	Excitation of Loop Oscillations			}

The exciter or trigger of the loop oscillations is very likely the
associated flare event to the east of the oscillating loop. If we
calculate the projected distance between the flare site
$(x_{flare}=+390\arcsec, y_{flare}=-410\arcsec)$ (section 2.2) and 
the apex of the oscillating loop
$(x_{apex}=+698\arcsec, y_{apex}=-243\arcsec, h_{apex} \approx 50\arcsec)$ 
(section 2.4), taking the stereoscopically triangulated 3D loop position
into account, we obtain and Euclidian distance of $L_{exc}=353\arcsec$
(256 Mm). Given the time delay between the start of the loop oscillation 
at the apex (19:12:12 UT) and the flare start (19:10:00 UT; $\pm 6$ s), 
we obtain the following 3D propagation speed of the exciter,
\begin{equation}
	v_{exc} = {L_{exc} \over T_{exc}}
		= {256,000\ {\rm km} \over 132\pm 6 {\ \rm s}} = 1940 \pm 125
		  \ {\rm km \ s}^{-1} \ ,
\end{equation}
which is a typical Alfv\'enic (magneto-acoustic) speed in the solar corona. 
Thus, we can conclude that the loop oscillation
is initially triggered by a fast MHD wave with Alfv\'enic speed. 
Moreover, the direction
of the initial excitation in west-ward direction is in the same
direction as the propagation direction of the fast MHD wave that
propagates with Alfv\'enic speed
concentrically away from the flare site. Although the
angle between the Alfv\'enic wave direction and the loop oscillation
amplitude cannot be determined with high accuracy, it is closer to
parallel than perpendicular, as would be expected in a vortex-shedding
scenario (Nakariakov et al.~2009), where the kink-mode oscillation occurs 
in perpendicular direction to the local plasma flow direction. 
In the case analyzed here, it appears that the plasma volume in
the westward direction is stretched out in the same direction,
probably following a narrow-angle cone of open magnetic field where the
CME escapes. An associated CME on the south-west side of the Sun
is visible in SOHO/LASCO and EIT movies. Generally, excitation of
kink-mode oscillations with vertical polarization are rare,
because they need special circumstances with
an exciter near the curvature center of the loop (Selwa et al.~2011).

\subsection{	External/Internal Density Ratio of Oscillating Loop}

Coronal seismology allows us to determine the mean magnetic field in a
loop in the kink-mode oscillation mode based on the internal Alfv\'en
speed $v_A$ inside the oscillating loop, 
\begin{equation}
	v_A = {B_i \over \sqrt{4 \pi \rho_i}} \ . 
\end{equation}
which can be related to the (fundamental) kink-mode period $P_{kink}$ using the
phase speed $c_k$ inside the flux tube (Roberts et al.~1984),
\begin{equation}
	P_{kink} = {2 L_{osc} \over c_k} 
	         = {2 L_{osc} \over v_A} \sqrt{{1 + \rho_e/\rho_i} \over 2}\ , 
\end{equation}
which depends on the total length $L_{osc}$ of the oscillating loop
and the densities externally ($n_e$) and internally ($n_i$) of the loop.

On the other hand, we can estimate the external Alfv\'en speed 
$v_{Ae}$, which depends on the external magnetic field $B_e$ and density $n_e$,
\begin{equation}
	v_{Ae} = {B_e \over \sqrt{4 \pi \rho_e}} \ . 
\end{equation}
If we interpret the exciter speed $v_{exc}$ between the flare site
and the (apex) location of the oscillating loop (Eq.~19) as an Alfv\'enic
wave, we obtain a direct measurement of the external Alfv\'en speed
(supposed the wave is not super-Alfv\'enic),
\begin{equation}	
	v_{Ae} = v_{exc} = {L_{exc} \over T_{exc}} \ .
\end{equation}
Moreover, since the magnetic pressure is generally dominant over thermal
pressure in the solar corona, the magnetic field $B_i$ internally and
$B_e$ externally to the loop boundary have to match for a self-consistent
magnetic field model of a loop embedded into an external plasma. 
Based on the definitions of the Alfv\'en speeds (Eqs.~20, 22), the ratio
of the Alfv\'en speeds depends then only on the density ratio,
\begin{equation}
	B_e = B_i \mapsto {v_{Ae} \over v_A} = \sqrt{ n_i \over n_e} \ ,
\end{equation}
which can be directly determined from the kink-mode period, loop length,
and exciter speed with Eqs.~(21) and (23),
\begin{equation}
	{n_i \over n_e} = {1 \over 2} 
	\left( {L_{exc} \over L_{osc}} {P_{kink} \over T_{exc}} \right)^2 - 1
	\ .
\end{equation}
Some uncertainty arises from the unknown
positions of the actual nodes of the kink mode. Since the oscillation
has only been observed over a loop segment of $L_{seg}=123$ Mm, which
is a lower limit of the node separation, while the stereoscopically
triangulated full loop length $L_{loop}$ down to the solar surface
represent an upper limit, we might estimate a realistic value with
uncertainty from the arithmetic mean of the two limits, i.e.,
$L_{osc} \approx (L_{loop} + L_{seg})/2 = 143 \pm 20$ Mm. 
Thus, for the physical parameters determined above 
($L_{exc}=256$ Mm, $T_{exc}=132\pm10$ s), and the best-fit values of
the primary oscillating loop (Fig.~8: 
$L_{osc} \lapprox (L_{loop}+L_{seg})/2=143\pm20$ Mm, $P_{kink}=375.6$ s), 
we obtain a value of ${n_i / n_e} \approx 12.0\pm1.8$, or an inverse ratio 
of ${n_e / n_i} \approx 0.08 \pm 0.01$. 
This value is commensurable with alternative methods, using stratified
hydrostatic density models of the background corona, for which a
statistical average of ${n_e / n_i} = 0.30 \pm 0.16$ was found.
(Aschwanden et al.~2003).
Consequently, for a density ratio of $n_e/n_i \approx 1/12$, the ratio of
the Alfv\'en speeds is then expected to be 
$(v_{Ae} / v_A) = \sqrt{ n_i / n_e} \approx \sqrt{12} \approx 3.4$.
In our case, the external Alfv\'en speed is 
$v_{Ae} \approx 1940\pm100$ km s$^{-1}$,
and the internal Alfv\'en speed is $v_A \approx 560 \pm 100$ km s$^{-1}$.

\subsection{	Magnetic Field Modeling 				}

With this novel method of measuring the density ratio ${n_e/n_i}$ from
Alfv\'enic propagation speeds external and internal to the fluxtube, 
the magnetic field in the oscillating flux tube and immediate surroundings 
is then fully constrained with Eqs.~20 and 21 (Nakariakov et al.~1999),
\begin{equation}
	B = {L_{osc} \over P_{kink}} 
	\sqrt{8 \pi \mu m_p n_i ( 1 + n_e/n_i) } \ ,
\end{equation}
for which we obtain $B = 4.0 \pm 0.7$ G, based on the measurements of 
$L_{osc}=(143\pm20) \times 10^8$ cm, 
$P_{kink}=375.6$ s, 
$n_i=(1.9 \pm 0.3) \times 10^8$ cm$^{-3}$,
and the density ratio $(n_e/n_i) \approx 0.08 \pm 0.01$.

This magnetic field scenario can be tested with observed magnetic field
data from the {\sl Helioseismic and Magnetic Imager (HMI)} on SDO. 
In Fig.~12 (bottom) we show a HMI magnetogram recorded at 19:04:16 UT,
at the beginning of the analyzed time interval. The flare location is
situated in the core of the AR, right at the neutral line with the largest
magnetic flux gradient, while the oscillating loop is located beyond the
western boundary of the active region in a low magnetic-field region that
is governed by a ``salt-and-pepper pattern'' of positive and negative
magnetic pores (see enlargement in Fig.~13 top left). A potential-field
source surface (PFSS) model calculation is shown in Fig.~12 (top panel),
which is dominated by a bipolar arcade above the neutral line in
east-west direction. 

The magnetic field in the environment of the oscillating loop can be
modeled with potential-field or non-potential field models, but both
are known to show misalignments with the 3D geometry of stereoscopically 
triangulated loops of the order of $\alpha_{mis} \approx 20^\circ-40^\circ$
(DeRosa et al.~2009; Sandman et al.~2009), while simple potential field 
models calculated from a small set of unipolar magnetic charges
(Aschwanden and Sandman 2010) or magnetic dipoles (Sandman and
Aschwanden 2010) achieved a reduced misalignment of $\alpha_{mis}
= 13^\circ-20^\circ$. For a simple plausibility test of the magnetic
field strength inferred from coronal seismology, we model the 3D field
at the location of the oscillating loop with an analytical model of two
unipolar charges with opposite magnetic polarities that are buried in 
depths $z_1$ and $z_2$ and have maximum longitudinal magnetic field strengths 
of $B_{\parallel 1}=+187$ G and $B_{\parallel 2}=-63$ G at the observed 
positions $(x_1,y_1)$ and $(x_2,y_2)$ of the nearest magnetic pores 
in the HMI magnetogram (marked with circles in Fig.~13, top left panel),
where $(x,y,z)$ is a cartesian coordinate system with the xy-plane
parallel to the solar surface. The corresponding absolute field strengths
vertically above the buried charges are 
$B_1=B_{\parallel 1}/\cos(\vartheta_1)=296$ G and 
$B_2=B_{\parallel 2}/\cos(\vartheta_2)=-89$ G,
where $\vartheta_j$ are the line-of-sight angles.
Thus, in this model we have only the two free variables
of the depths $z_1$ and $z_2$ to fit the model of resulting magnetic
field lines to the observed loop. The magnetic field resulting from
the superposition of two unipolar magnetic charges is then given by
(Aschwanden and Sandman 2010),
\begin{equation}
        {\bf B}({\bf x}) = \sum_{j=1}^N {\bf B}_j({\bf x})
        = \sum_{j=1}^N  B_j
        \left({z_j \over r_j}\right)^2 {{\bf r}_j \over r_j} \ ,
\end{equation}
in terms of the vector ${\bf r}_j = [(x-x_j), (y-y_j), (z-z_j)]$,
with ${\bf x}_j=(x_j, y_j, z_j)$ being the locations of the buried
unipolar magnetic charges and $B_j$ the magnetic field strength
at the solar surface above the magnetic charges. 
The ratio of the two free variables $z_1$ and $z_2$ determine the
asymmetry of the field lines. For the observed oscillating loop
we find values of $z_1=0.5\arcsec$ and $z_2=1.5\arcsec$ to reproduce
approximately the observed shape (Fig.~13, bottom left). The field
line that closest fits the projected location of the oscillating
loops has magnetic field strenghts of $B_1=296$ G and $B_2=29$ G
at the photospheric field line footpoints and $B=6$ G at the apex, which
compares favorably with the magnetic field strength of 
$B_{kink}=4.0\pm0.7$ G deduced from coronal seismology. 

However, since the magnetic field $B(s)$ varies along the loop,
the Alfv\'en speed varies proportionally and the
Alfv\'enic transit time during one oscillation period is given by
\begin{equation}
	P = \int_0^P dt = \int_0^{2L} {1 \over v_A(s)} ds \ ,
\end{equation}
which defines an average magnetic field $\langle B \rangle$ that is equivalent
to a fluxtube with the same kink-mode period $P$ and a constant 
magnetic field value $\langle B \rangle$ by
\begin{equation}	
	\langle B \rangle = \left[ \int B(s)^{-1} ds \right]^{-1} \ ,
\end{equation}
for which we obtain $\langle B \rangle = 11$ G, which is a factor of 1.8 higher
than the minimum value at the apex, $B_{apex}=6$ G, or a factor of 2.8 higher
than inferred from seismology, $B_{kink}=4.0\pm0.7$ G.  
This difference between the seismological and magnetogram-constrained
magnetic field value, derived for the first time for an oscillating
loop here to our knowledge, is perhaps not too surprising, given the
ambiguity of potential-field models, non-potential field models,
and uncertainties of the footpoint locations (which require
stereoscopic information). 

\section{	DISCUSSION				}

In this well-observed loop oscillation event, which we analyzed with
AIA/SDO, HMI/SDO, and EUVI/ STEREO, we derived a comprehensive number of
physical parameters (listed in Table 1) that could not be determined 
to such a degree in previous observations. In the following discussion
we compare the observational results with theoretical models, 
predictions, and discuss interpretational issues. 

\subsection{	Coupled Kink-Mode and Cross-Sectional Oscillations      }

The basic theory for fast magneto-acoustic waves, which predicts kink 
and sausage eigen-modes for slow (acoustic) and fast (Alf\'enic)
MHD waves, has been derived for a straight (slender)
cylindrical fluxtube (e.g., Edwin and Roberts 1983). For such an
idealized geometry, the periods of the fast kink and sausage mode have
quite different regimes, and the sausage mode has a wavenumber cutoff
with no solution of the dispersion relation for $ka \lapprox 1$
(with $k$ the wave number and $a$ the fluxtube radius), which 
corresponds to a cutoff at a phase speed of $v_{ph}=v_{Ae}$.
From this theory, no sausage eigen-mode is predicted for periods
that correspond to kink-mode oscillations, $P_{kink}=2L/v_A$.
In contrast, our analysis clearly demonstrates the presence of a kink-mode 
with coupled sausage-like behavior, as measured by the cross-sectional 
loop width variations and anti-correlated density variations.
The question arises why this dynamical behavior is not predicted
by existing theory? One possible explanation is that the loop length
is not constant but changes as a function of time in synchronization
with the transverse oscillation amplitude. This is most plausibly seen
in Fig.~1, where the excitation direction originating from the flare
location propagates in approximately the same direction as the loop plane, 
and thus excites a significant component of the ``vertical'' polarization mode 
(i.e., the loop plane and the oscillation plane are near-parallel), 
as inferred for one case in Wang and Solanki (2004) and analytically studied 
in Verwichte et al.~(2006a,b). Most kink-mode oscillations have a horizontal
polarization, as determined with STEREO (e.g., Verwichte et al.~2009),
but density oscillations have alse been noted in previous kink-mode
oscillations (e.g., Verwichte et al.~2009, 2010).
If the loop oscillates in vertical
polarization, the length of the loop can vary during the kink-mode
oscillations, with a linear dependence on the oscillation amplitude
to first order (in the elliptical approximation, see Eqs.~9 and 10). Thus,
the periodic shrinking and stretching of the fluxtube is likely to
cause a bulging and thinning of the central loop cross-section, which
is exactly what a sausage mode does. A consequence of the length variation
$L(t)$ is also a magnetic field variation $B(t)$, which scales reciprocally
to the cross-sectional area of the sausage mode, i.e., $B(t) \propto A^{-1}(t)$,
due to the conservation of the magnetic flux, i.e., $B(t)A(t)$=constant. 

The coupling of kink-mode and (sausage-like) cross-section and density 
oscillations thus might be a special case that occurs only when the
loop length is varied, which most likely occurs for vertical polarization
and requires an initial excitation in direction of the loop plane.
It would be interesting to investigate this prediction of coupled
cross-section and density oscillations 
as a function of the exciter direction or kink-mode polarization,
which depends on the location and orientation of the loop plane with
respect to the propagation direction of a flare or CME-related
disturbance. Since CME bubbles and erupting flux ropes get stretched
out during the initial expansion, it is natural that ambient magnetic
field lines become stretched too, which applies also to oscillating loops.
Statistics on different polarization types of kink-mode oscillations
is still small and their identification based on difference images
is often ambiguous (Wang et al.~2008). 

An alternative interpretation of the amplitude-correlated flux variation
is an aspect-angle change of the
oscillating loop, which causes a variable line-of-sight column depth of
the loop diameter $w(t) = w_0 \cos[\vartheta(t)]$ (Cooper et al.~2003), 
and hence would introduce a variation of the optically-thin EUV flux 
$f(t) \propto n_e^2(t) w(t)$. However, the observed flux variation with a
mean of $\approx 24\%$ would require an aspect angle change of 
$\Delta \vartheta \approx 40^\circ$, which is inconsistent with the 
observed stationarity of the loop shape during the entire oscillation 
episode. 

\subsection{	Multi-Loop Oscillations 		}

Evidence that multiple loops or strands are involved in this oscillation
event is shown in Fig.~8, where we found slightly different periods 
(by $\approx 10\%$), amplitudes, centroid positions, and possibly different
lengths (although not directly measured). The eigen-modes
in a two-slab system was studied in Arregui et al.~(2008) and it was
found that the kink-mode periods may differ from a single loop
when the distance between the loops is less than a few loop diameters.
In our case, the projected centroid position is displaced by $\Delta x=0.7$ Mm,
while the loop diameters are $w \approx 4.9\pm0.6$ Mm, so they could be
close to each other. Luna et al.~(2008) simulated numerically the
MHD behavior of two parallel loops and found four collective modes, kink 
(asymmetric) and sausage (symmetric) modes in both parallel and
perpendicular direction to the plane that contains the axis of both loops,
with four different frequencies, which is a generalization of the two modes
of a single-loop oscillation. However, analytical solutions of a two-loop 
system yields only two different frequencies (Van Doorsselaere et al.~2008), 
which might differ from the numerical results of four different frequencies
(Luna et al.~2008) due to the neglect of higher-order terms (Ruderman and 
Erdelyi 2009).

A multi-threaded model with four loop threads was
modeled with a 3D MHD code (Ofman 2009). For parallel threads, the
evolution of the ensemble exhibits the same period and damping rate
as a single loop, but for twisted threads, the periods become irregular
and the damping much stronger, which seems not to apply to our case here. 
Either the multiple loops are near-parallel or sufficiently distant to
each other. 

Resonant absorption in complicated multi-strand loops was investigated
by Terradas et al.~(2008) and it was found that the damping behavior
is not compromised by the complicated geometry of composite loops.
One theoretical prediction of multi-loop oscillations is that
the collective width $w(t)$ increases with time due to a
shear instability (Terradas 2009), but we do not observe such an effect
(Fig.~11, bottom right panel), either because the two oscillating loops 
are not in sufficiently close spatial proximity or because the lifetime 
of the oscillating loops in the detected wavelength is too short. 

\subsection{	Damping by Resonant Absorption 		}

An unusual property of this oscillation event is that we do not observe
any significant damping of the kink-mode amplitude over the duration 
of the oscillatory episode, so the ratio of the damping time to the period
must be much longer than the observed number of periods, 
i.e., $\tau_D/P \gg 4$. This is in contrast to a statistical
sample of 11 well-observed events with TRACE, where strong 
damping was found to be the rule, i.e., with $\tau_D/P \approx 1.8 \pm 0.8$ 
(Aschwanden et al.~2002).  

Resonant absorption as a damping mechanism for kink-mode oscillations was
considered in Goossens et al.~(2002). The ratio of the damping time to
the period was calculated for resonant absorption by Ruderman and Roberts
(2002) for a thin-boundary layer and by Van Doorsselaere et al.~(2004)
for thick boundaries, 
\begin{equation}
	\left( {\tau_D \over P}_{thin} \right) = q_{TB} {2 \over \pi}
	\left( {r_{loop} \over l_{skin} } \right)
	{(1+q_n) \over (1 - q_n)} \ ,
\end{equation}
where $q_{TB}\approx 0.75$ is the correction factor for the thick-boundary
layer, $l_{skin}$ is the skin depth or thickness of the loop boundary
that contains a density gradient, and $q_n=n_e/n_i$ is the ratio of
the external to the internal electron density in the loop. 
This density ratio was previously measured to $q_n = 0.30\pm0.16$, based on
loop flux intensities and hydrostatic models of the background corona
(Aschwanden et al.~2003), and a skin depth ratio of 
$r_{loop}/l_{skin}=1.5\pm0.2$ was inferred, and hence the typical ratio of the
damping time to the oscillation period was found to be $\tau_D/P \approx 1.3$. 

In our case, a similar density ratio of $q_n \approx 0.08$ was measured.
We can reconcile the observed long damping time ratio of $\tau_D/P \gg 4$
only with a very small skin depth of $l_{skin}/r_{loop} \ll 1/4$. 
While previously analyzed kink-mode oscillations with TRACE exhibited
typical temperatures of $T_e \approx 1.0-1.5$ MK, we deal here with a
significantly cooler loop with a temperature of $T_e \approx 0.5$ MK.
It appears that such cooler loops have either a smaller skin depth
or larger loop diameters than the warmer coronal loops, but no hydrodynamic 
model is known that predicts such an effect. 

\subsection{	Magnetic Field Comparisons		}

Coronal seismology determines the magnetic field strength by setting
the kink-mode period $P_{kink}$ equal to the Alfv\'enic crossing time 
$2L/v_A$ forth and back along the loop length $L$, which yields
a relationship for the magnetic field $B_{kink}$ as a function of the
loop length $L$, period $P_{kink}$, internal $n_i$ and external
density $n_e$ (Eqs.~21 and 26). This method is one of the foundations
of coronal seismology, initially applied by Roberts et al.~(1984),
Aschwanden et al.~(1999), and Nakariakov and Ofman (2001). 
In principle, this analytical relationship can be put to the test by
3D MHD simulations of kink-mode oscillations of a plasma fluxtube
by comparing the theoretical with the experimental values of the
kink-mode oscillation periods $P_{kink}$ or magnetic fields $B$. 
Such a test was conducted by DeMoortel and Pascoe (2009), but
surprisingly the coronal seismology formula predicted a field strength
($B_{kink}=15-30$ G) that was about a factor of 1.5 higher than the 
input values of $B=10-20$ G of the MHD simulation.  

Here we attempted to validate the seismological magnetic field value
($B_{kink}=4.0 \pm 0.7$ G) by a potential-field model that consists of
two unipolar magnetic charges with opposite polarities that are buried 
near the footpoints of the oscillating loop and are constrained by
the longitudinal magnetic field strengths observed in HMI magnetograms.
The best-fit field line yielded a magnetic field value of
$B_{apex}=6$ G, which is a factor 1.4 higher than the seismological value.
If we correct for the variable Alfv\'en speed along the loop, we predict
a seismological value of $B_{avg}=11$ G, which is a factor of 2.8 higher
than the theoretical value. We note that the discrepancy of the
our best-fit potential field model is in opposite direction to the
discrepancy found from 3D MHD simulations by DeMoortel and Pascoe (2009).
We believe that the discrepancy from magnetic field modeling methods
mostly stems from the uncertainty of the footpoint locations, the
spatial resolution of magnetograms, and the ambiguity of potential
and non-potential field models. Stereoscopically triangulated loop
oscillations hold the promise for obtaining more accurate measurements
of the loop length and footpoint location. The most powerful 
self-consistency test needs to employ a combination of stereoscopy, 
numerical 3D MHD simulations, coronal seismology theory, and analytical 
magnetic field models. 

\section{	CONCLUSIONS				}

Here we present the first analysis of a loop oscillation event observed
with AIA/SDO, which occurred on 2010-Oct-16, 19:05-19:30 UT. 
The capabilities of AIA enable us for the first time to study such an
event with sufficiently high cadence, spatial resolution, and comprehensive
temperature coverage, which enables us to derive all important physical
parameters. In addition, magnetic modeling with HMI data can validate the
magnetic field measurements based on coronal seismology. The major 
observational findings, interpretations, and conclusions are:

\begin{enumerate}
\item{A flare with an associated CME that escapes the Sun along a narrow
cone (in westward direction) excites kink-mode oscillations with a period 
of $P=6.3$ min in a loop at a distance of $L_{exc}=256$ Mm away from the 
flare site, after a time delay of $T_{exc}=132$ s, which yields an exciter 
speed of $v_{exc}=L_{exc}/T_{exc} \approx 1900$ km s$^{-1}$, which we
interpret as a magneto-acoustic wave with Alfv\'enic speed and can be used
as a direct measurement of the average external Alfv\'en speed 
$v_{Ae}=v_{exc}$ outside the oscillating loop.}

\item{The direction of the excitation and kink-mode oscillation amplitude
is about in the same direction as the loop plane, which corresponds to
a vertical polarization of the kink mode, causing a periodic stretching 
of the loop length and coupled cross-section and density oscillations,
evident from the compression and rarefaction of the
density, which produces an intensity variation that is amplified with the
fourth power of the amplitude displacement. This behavior of kink modes
with coupled cross-sectional and density variations 
are unusual and perhaps occur only in vertical polarization.
They are not predicted by theory, which needs to be generalized
for temporal variations of the loop length $L(t)$.}

\item{There is evidence for a multi-loop system that is involved in the
coupled kink and cross-sectional oscillations, consisting of at least two
loop strands that have slightly different periods ($\approx 10\%$) but
are excited in phase at the beginning. The fact that the two major
oscillating loop strands are not synchronized to the same period
indicates a spatial separation of more than a few loop diameters.}

\item{A full DEM analysis with all 6 coronal AIA temperature filters
yields a temperature of $T \approx 0.5$ MK and a density of $n_e \approx
2 \times 10^8$ cm$^{-3}$. Consequently, the loop oscillations are primarily
observable in the 171 \ang\ filter, very faint in the 131 and 193 \ang\
filter, and essentially undetectable in the other filters. From this
temperature and density measurement we estimate a radiative cooling
time of $\tau_{rad}=46$ min, which explains the loop lifetime of 
$\tau_{life}=27$ min in the 171 \ang\ filter.}

\item{The measurement of the external Alfv\'en speed $v_{Ae} \approx 1900$
km s$^{-1}$ from the exciter speed and the internal Alfv\'en speed
$v_A = 560$ km s$^{-1}$ from the kink-mode period provides a direct
measurement of the density ratio external and internal to the loop,
$n_e/n_i=0.08\pm0.01$, which is commensurable with earlier hydrostatic
models of the background corona ($n_e/n_i=0.30\pm0.16$; Aschwanden 
et al.~2003). This value provides a fully constrained magnetic field
measurement of the oscillating loop by coronal seismology,
$B_{kink}=4.0 \pm 0.7$ G.}

\item{For an independent estimate of the magnetic field in the oscillating
loop we used a potential-field model with two unipolar magnetic charges,
constrained by the photospheric magnetic field strengths 
($B_1=+296$ G, $B_2=-89$ G) obtained from HMI/SDO magnetograms near the 
footpoints of the oscillating loop, which were localized by
stereoscopic triangulation from STEREO/EUVI-A images. A best-fit model
yields a magnetic field strength of $B_{apex}=6$ G at the loop apex,
or $B_{avg}=11$ G when averaged along the loop. This independent test
validates the coronal seismological value within a factor of $\approx 2$.}

\item{The oscillating loop exhibits no detectable damping over the
observed four periods, which is unusual, compared with the statistical
values of $\tau_D/P = 1.8 \pm 0.8$ found from previous measurements.
Damping by resonant absorption can only be reconciled with this observation
if the skin layer (of the density gradient at the loop boundary) is much
smaller than the loop radius. It is not clear if this property is a
consequence of the unusual low loop temperature of $T \approx 0.5$ MK}.
\end{enumerate}

The excellent quality of the AIA data have provided more physical
parameters of a coronal loop oscillation event than it was possible
to determine in previous TRACE observations, especially due to the
much better cadence of 12 s, which allows us also to resolve
multi-loop oscillations, spatially and temporally. The measurements
of more physical parameters provide stronger constraints on the theory
and raise new problems that need to be addressed by analytical theory 
or MHD simulations: (1) What is the 3D geometry and timing of the exciter
mechanism and how does it affect the polarization of kink-mode 
oscillations?  (2) Can we explain the coupling of kink mode and (sausage-like)
cross-sectional and density oscillations?
(3) Can we explain kink-mode oscillations with no damping?
(4) How do multi-loop oscillations interact with each other and
how do the MHD wave modes couple? (5) How accurate are magnetic field 
measurements based on coronal seismology and how can they be validated
with magnetic field models? Progress in these questions calls for
modeling that combines stereoscopy, numerical 3D MHD simulations, 
coronal seismology theory, and analytical magnetic field models. 

\acknowledgements {\sl Acknowledgements:} 
We acknowledge constructive and helpful discussions with Valery Nakariakov,
Erwin Verwichte, Jaume Terradas, Robertus Erdelyi, Richard Morton,
Leon Ofman, Tom Van Doorsselaere, Ineke De Moortel, Mag Selwa, 
Michael Ruderman, and James McAteer, mostly during a meeting on 
``Coronal Heating and Waves''
sponsored by the Royal Society in London, January 5-7, 2011.
This work is partially supported by NASA contract NAG5-13490, 
NASA contract NNG04EA00C of the SDO/AIA instrument, and
NRL contract N00173-02-C-2035 of the NASA STEREO mission. 


\section*{REFERENCES}

\def\ref#1{\par\noindent\hangindent1cm {#1}}
\def\aap {{\sl Astron. Astrophys.}\ } 
\def\apj {{\sl Astrophys. J.}\ } 
\def\grl {{\sl Geophys. Research Lett.}\ } 
\def\sp  {{\sl Solar Phys.}\ } 
\def\ssr {{\sl Space Science Rev.}\ } 

\ref{Andries, J., van Doorsselaere, T., Robedrts, B., Verth, G.,
	Verwichte, E., and Erdelyi, R. 2009, \ssr 149, 3.}
\ref{Arregui, I., Terradas, J., Oliver, R., and Ballester, J.L.
	2008, \apj 674, 1179.}
\ref{Asai, A., Shimojo, M., Isobe, H., Morimoto, T., Yokoyama, T., 
	Shibasaki, K., and Nakajima, H. 2001, \apj 562, L103.}
\ref{Aschwanden, M.J., Fletcher, L., Schrijver, C., and Alexander, D.
 	1999, \apj 520, 880.}
\ref{Aschwanden, M.J., DePontieu, B., Schrijver, C.J., and Title, A.
 	2002, \sp 206, 99.}
\ref{Aschwanden, M.J., Nightingale, R.W., Andries, J., Goossens, M.,
	and Van Doorsselaere, T. 2003, \apj 598, 1375.}
\ref{Aschwanden, M.J. 2004, 
	{\sl Physics of the Solar Corona - An Introduction},
	Springer and Praxis, New York, 892p.}
\ref{Aschwanden, M.J., Nakariakov, V.M., and Melnikov, V.F. 2004,
	ApJ 600, 458.}
\ref{Aschwanden, M.J. and Nightingale, R.W. 2005, \apj 633, 499.}
\ref{Aschwanden, M.J. 2006, Phil. Trans. Royal Society, 
	in "MHD Waves and Oscillations in the Solar Plasma", 
	(eds. R. Erdelyi, and J.M.T. Thompson), Vol. 364, p.417.}
\ref{Aschwanden, M.J. and Terradas, J. 2008, \apj 686, L127.}
\ref{Aschwanden, M.J. and Sandman, A.W. 2010, Astron. J. 140, 723.}
\ref{Aschwanden, M.J. and Boerner, P. 2011, \apj (subm).}
\ref{Banerjee, D., Erdelyi, R., Oliver, R., O'Shea, E. 2007,
	\sp 246, 3.}
\ref{Boerner, P., Edwards, C., Lemen, J., Rausch, A., Schrijver, C.,
        Shine, R., Shing, L., Stern, R., Tarbell, T., Title, A.,
        and Wolfson, C.J. 2011, {\sl Initial calibration of the Atmospheric
        Imaging Assembly Instrument}, (in preparation).}
\ref{Cooper, F.C., Nakariakov, V.M., and Tsiklauri, D. 2003, \aap 397, 765.}
\ref{DeForest, C.E., and Gurman,J.B. 1998, ApJ 501, L217.}
\ref{DeMoortel, I., Ireland, J., Walsh, R.W., and Hood, A.W. 2002a, 
	\sp 209, 61.}
\ref{DeMoortel, I., Hood, A.W., Ireland, J., and Walsh, R.W.
 	2002b, \sp 209, 89.}
\ref{DeMoortel, I. and Pascoe, D.J. 2009, \apj 699, L72.}
\ref{DeRosa, M.L., Schrijver, C.J., Barnes, G., Leka, K.D., Lites, B.W., 
	Aschwanden, M.J., Amari, T., Canou, A., McTiernan, J.M., Regnier, S., 
	Thalmann, J., Valori, G., Wheatland, M.S., Wiegelmann, T., 
	Cheung, M.C.M., Conlon, P.A., Fuhrmann, M., Inhester,B., and 
	Tadesse,T. 2009, \apj 696, 1780.}
\ref{Edwin, P.M. and Roberts, B. 1983, \sp 88, 179.}
\ref{Erdelyi, R., Petrovay, K., Roberts, B., and Aschwanden, M.J. (eds.)
	2003, "Turbulence, Waves and Instabilities in the Solar Plasma",
	NATO Series II, Kluwer Academic Publishers, Dordrecht.} 
\ref{Erdelyi, R. and Taroyan, Y. 2008, \aap 489, L49.}
\ref{Goossens, M., Andries, J., and Aschwanden, M.J. 2002,
	\aap 394, L39.}
\ref{Katsiyannis, A.C., Williams, D. R., McAteer, R.T.J., Gallagher, P.T., 
	Keenan, F.P., and Murtagh, F. 2003, \aap 406, 709.}
\ref{Kliem, B., Dammasch, I.E., Curdt, W., and Wilhelm, K. 2002,
	\apj 568, L61.}
\ref{Lemen, J. and AIA Team, 2011, \sp (in preparation).}
\ref{Luna, M., Terradas, J., Oliver, R., and Ballester, J.L. 2008,
	\aap 457, 1071.} 
\ref{Melnikov, V.F., Shibasaki, K., and Reznikova, V.E. 2002, \apj 580, L185.}
\ref{Morton, R.J. and Erdelyi, R. 2009, \apj 707, 750.}
\ref{Morton, R.J. and Erdelyi, R. 2010, \aap 519, A43.}
\ref{Morton, R.J., Erdelyi, R., Jess, D.B., and Mathioudakis, M. 2011,
	\apj 728, L1.}
\ref{Nakariakov, V.M., Ofman,L., DeLuca,E., Roberts,B., and Davila,J.M.
 	1999, Science 285, 862.}
\ref{Nakariakov, V.M. and Ofman, L. 2001, \aap 372, L53.}
\ref{Nakariakov, V.M. and Verwichte, E. 2005, Living Reviews in Solar 
	Physics, 2, 3.}
\ref{Nakariakov, V.M., Aschwanden, M.J., and Van Doorsselaere, T. 2009,
	\aap 502, 661.}
\ref{Ofman, L., Romoli, M., Poletto, G., Noci, G., and Kohl, J.K.
 	1997, \apj 491, L111.}
\ref{Ofman, L. 2009, \apj 694, 502.}
\ref{Press, W.H., Flannery, B.P., Teukolsky, S.A., and Vetterling, W.T.
 	1986, {\sl Numerical recipes, The Art of Scientific Computing},
 	Cambridge University Press: Cambridge.}
\ref{Roberts, B., Edwin, P.M., and Benz, A.O. 1984, \apj 279, 857.}
\ref{Roberts, B. and Nakariakov, V.M. (2003), in
	"Turbulence, Waves and Instabilities in the Solar Plasma",
	(Eds. Erdelyi, R., Petrovay, K., Roberts, B., and Aschwanden, M.J.)
	NATO Series II, Kluwer Academic Publishers, Dordrecht, p. 167.} 
\ref{Roberts, B. 2004, in "Waves, Oscillations, and small scale events
	in the solar atmosphere", (ed. H. Lacoste)m ESA, ESTEC Noordwijk,
	ESA SP-547, 1.} 
\ref{Ruderman, M.S. and Roberts, B. 2002, \apj 577, 475.}
\ref{Ruderman, M.S. and Erdelyi, R. 2009, \ssr 149, 199}.
\ref{Sandman, A., Aschwanden, M.J., DeRosa, M., Wuelser, J.P. and Alexander,D.
 	2009, \sp 259, 1.}
\ref{Sandman, A.W. and Aschwanden, M.J. 2011, \sp (in press).}
\ref{Selwa, M., Solanki, S.K., and Ofman, L. 2011,
	{\sl The role of active region loop geometry - II. Symmetry
	breaking in 3D active region: Why are vertical kink oscillations
	observed so rarely?}, \apj (in press).}
\ref{Taroyan, Y. and Erdelyi, R. 2009, \ssr 149, 229.}
\ref{Terradas, J., Arregui, I., Oliver, R., Ballester, J.L., 
	Andries, J., and Goossens, M. 2008, \apj 679, 1611.}
\ref{Terradas, J. 2009, SSRv 149, 255.}
\ref{Tomczyk, S., McIntosh, S.W., Keil, S.L., Judge, P.G., Schad, T., 
	Seeley, D.H., and Edmondson, J. 2007, Nature 317, 1192.}
\ref{Thompson, B.J., Plunkett, S.P., Gurman, J.B., Newmark, J.S.,
	St.Cyr, O.C., and Michels, D.J. 1998, \grl 25, 2465.}
\ref{Thompson, B.J., Gurman, J.B., Neupert, W.M., Newmark, J.S., 
	Delaboudiniere, J.P. St.Cyr, O.C., Stezelberger, S., Dere,K.P., 
	Howard,R.A., and Michels,D.J. 1999, ApJ 517, L151.}
\ref{Van Doorsselaere, T., Andries, J., Poedts, S., and Goossens, M.
 	2004, \apj 606, 1223.}
\ref{Van Doorsselaere, T., Ruderman, M.S., and Roberstson, D. 2008,
	\aap 485, 849.} 
\ref{Verwichte, E., Nakariakov, V.M., Ofman, L., and DeLuca, E.E.
 	2004, \sp 223, 77.}
\ref{Verwichte, E., Foullon, C., and Nakariakov, V.M. 2006a, \aap 446, 1139.}
\ref{Verwichte, E., Foullon, C., and Nakariakov, V.M. 2006b, \aap 449, 769.}
\ref{Verwichte, E., Aschwanden, M.J., Van Doorsselaere, T., Foullon, C.,
	and Nakariakov, V.M. 2009, \apj 698, 397.}
\ref{Verwichte, E., Foullon, C., and VanDoorsselaere, T. 2010, 
 	\apj 717, 458.}
\ref{Wang, T.J., Solanki, S.K., Curdt, W., Innes, D.E., and Dammasch, I.E.
 	2002, ApJ 574, L101.}
\ref{Wang, T.J. and Solanki, S.K. 2004, \aap 421, L33.}
\ref{Wang, T.J. 2004, in "Waves, Oscillations, and small scale events
	in the solar atmosphere", (ed. H. Lacoste)m ESA, ESTEC Noordwijk,
	ESA SP-547, 417.} 
\ref{Wang, T.J., Solanki, S.K., and Selwa,M. 2008, \aap 489, 1307.}
\ref{Williams, D.R., Phillips, K.J.H., Rudawy, P., Mathioudakis, M., 
	Gallagher, P.T., O'Shea, E., Keenan, F.P., Read, P., and Rompolt, B.
 	2001, MNRAS 326, 428.}

\clearpage

\begin{deluxetable}{ll}
\tabletypesize{\normalsize}
\tablecaption{Observables and physical parameters of analyzed loop oscillation
event}
\tablewidth{0pt}
\tablehead{
\colhead{Parameter}&
\colhead{Value}}
\startdata
Date of observations  			& 2010-Oct-16 			\\
Time interval of analyzed observations  & 19:05-19:35 UT		\\
Time range of GOES flare 		& 19:07-19:12 UT		\\
Flare onset of impulsive phase 		& 19:10:00 ($\pm 6$ s) UT	\\
Start of loop oscillations		& 19:12:12 ($\pm 6$ s) UT 	\\
GOES flare class			& M2.9				\\
Active region number			& NOAA 1112			\\
Flare location				& [390\arcsec,-410\arcsec], W26/S20 \\
Location of oscillating loop footpoints & [685\arcsec,-305\arcsec],
                                          [615\arcsec,-268\arcsec] 	\\
Location of loop apex                   & [698\arcsec,-243\arcsec]      \\
Distance of flare to loop apex $L_{exc}$& 275 Mm			\\
Delay of flare start to loop oscillation $T_{exc}$ & $132\pm10$ s 	\\
Exciter speed $v_{exc} = v_{Ae}$ 	& $1940\pm 100$ km s$^{-1}$ 	\\
Height of loop apex $h_{apex}$	        & 37 Mm 			\\
Distance from Sun center		& $740\arcsec$ ($0.77 R_{\odot}$) \\
Full loop length $L_{loop}$		& 163 Mm			\\
Length of oscillating loop segment $L_{seg}$ & 123 Mm			\\
Loop curvature radius $r_{loop}$ 	& 52 Mm				\\
Loop FWHM diameter $w$			& $4.9\pm0.6$ Mm		\\
Loop inclination angle to vertical $\vartheta$ & $20^\circ\pm20^\circ$	\\	
Polarization angle of kink oscillation	& $\approx$ vertical		\\
Drift velocity of loop centroid $ds/dt$& 0.8 km/s (towards west)	\\
Oscillation period of loop $P$		& $375.6$ s ($6.3$ min)         \\
Oscillation amplitude of loop $a_1$	& $1.7\pm0.4$ Mm		\\
Number of oscillation periods $N_P$	& 3.6				\\
Loop lifetime $\tau_{life}=f_0/(df/dt)$ & 1650 s (27 min)		\\
Ratio of loop amplitude to radius $a_1^{max}/r_{loop}$ & 0.042		\\
Observed flux modulation $f_1/f_0)$     & 0.24 (0.18 predicted)		\\
Electron temperature $T_e$		& $0.57\pm0.14$ MK		\\
Temperature width $\sigma_{log(T)}$	& $0.18\pm0.10$			\\
Electron density $n_e$			& $(1.9\pm0.3) \times 10^8$ cm$^{-3}$\\
External Alfv\'en speed $v_{exc}=v_{Ae}$& $1940\pm 100$ km s$^{-1}$ 	\\
Internal Alfv\'en speed $v_{A}$ 	& $560\pm 100$ km s$^{-1}$ 	\\
External/internal density ratio $n_e/n_i$& $0.08\pm0.01$		\\
Magnetic field at loop apex $B_{apex}$  & $4.0\pm0.7$ G			\\
Magnetic field at loop footpoints $B_{foot}$ & $+296, -89$ G		\\
Damping time ratio $\tau_{damp}/P$      & $\gg 4$			\\
\enddata
\end{deluxetable}


\begin{figure}
\centerline{\includegraphics[width=\textwidth]{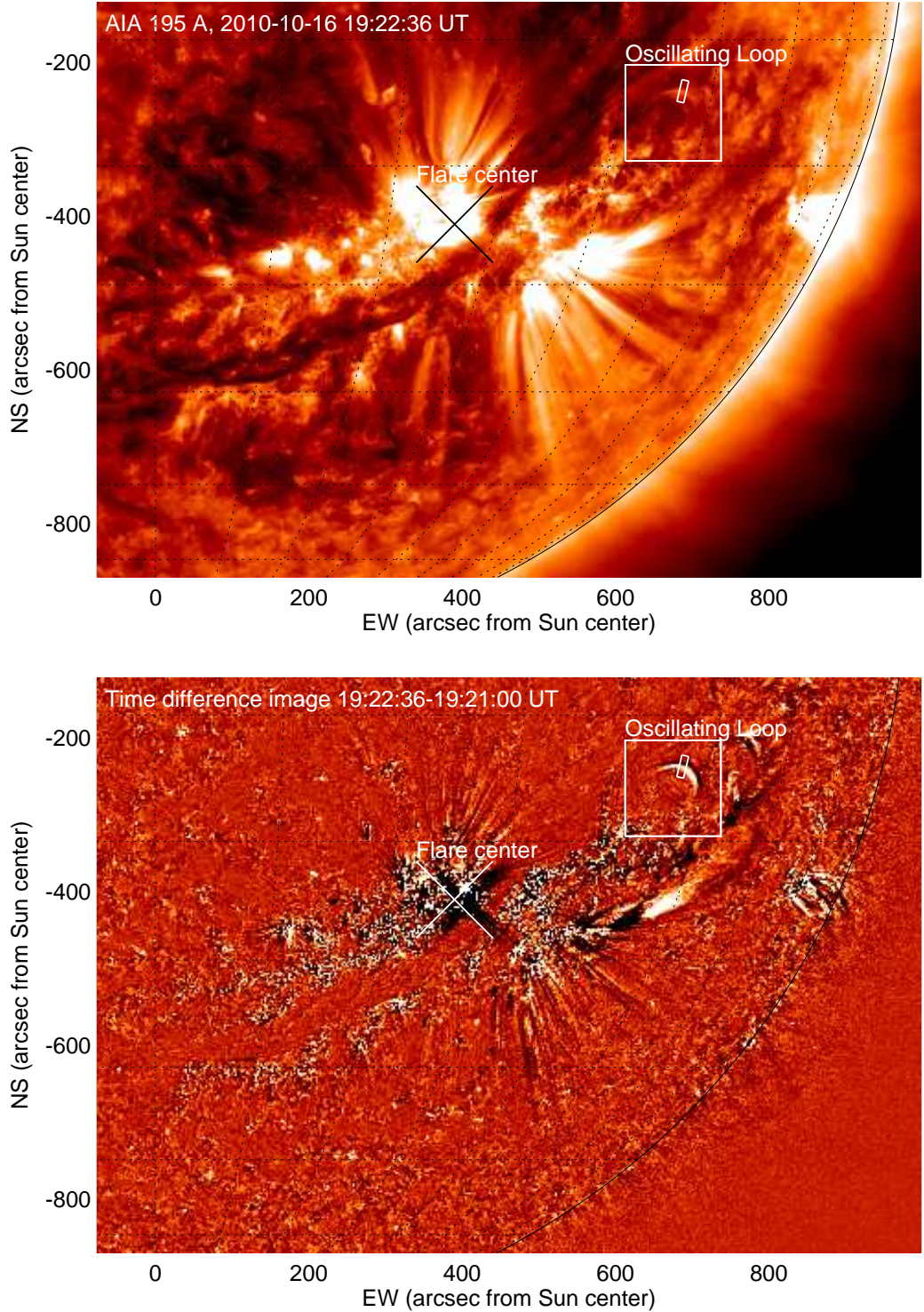}}
\caption{AIA 171 A image of flare observed on 2010-Oct-16 19:22:36 UT shown
with the flux on a logarithmic flux scale (top panel) and as difference image 
with respect to 19:21:00 UT (bottom panel). The flare location is marked with
a cross (in the center of the diffraction pattern) and a box indicates the
location of the oscillating loop. [See also movies in 171 \ang\
intensity and running-difference format that are available as supplementary
data in the electronic version of this journal].}
\end{figure}

\begin{figure}
\centerline{\includegraphics[width=0.9\textwidth]{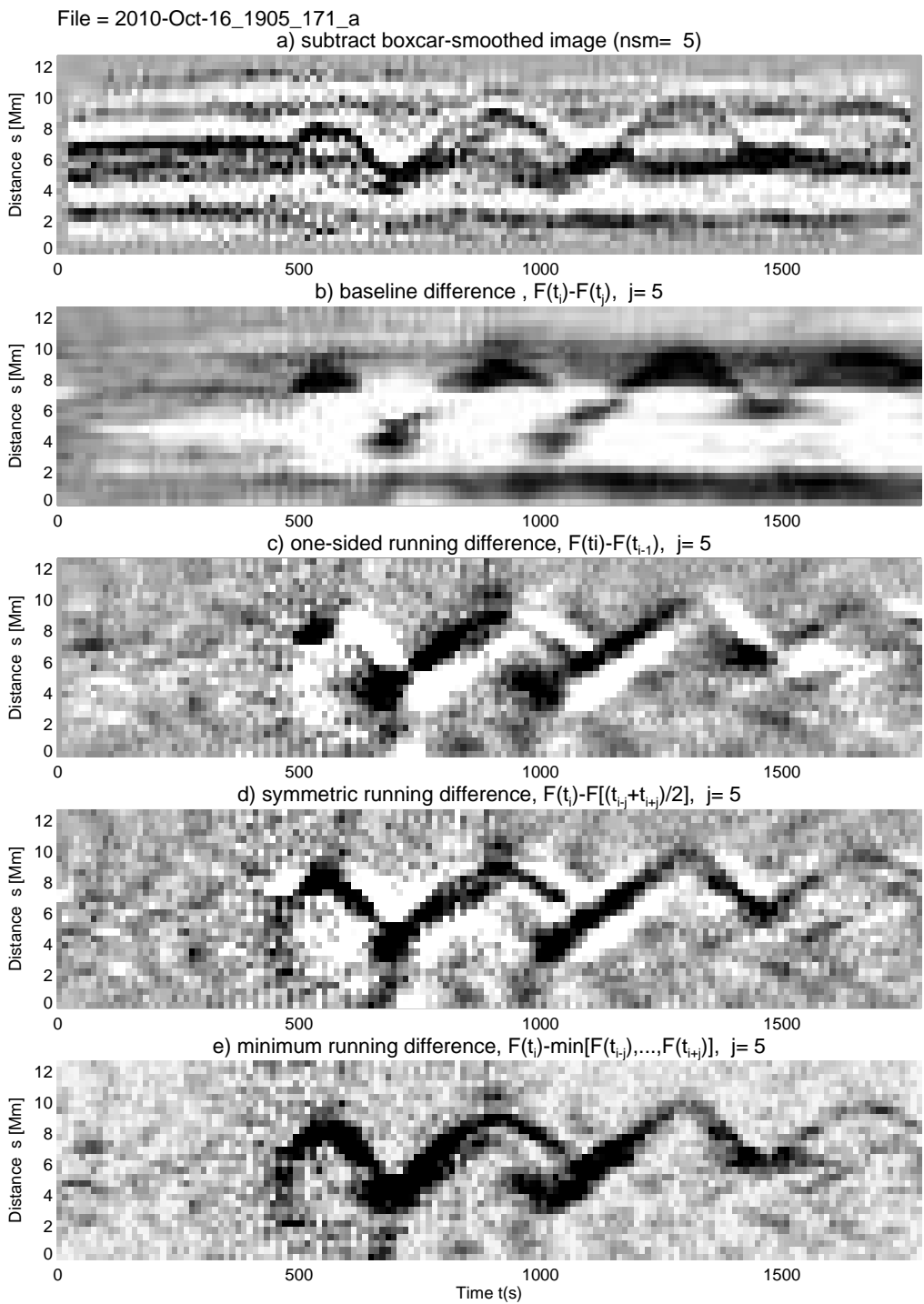}}
\caption{Five different enhancement algorithms to visualize the loop oscillation
in a time-slice plot with the time running along the x-axis 
(time range is 2010-Oct-16 19:05 --- 19:35 UT) and the spatial
coordinate (defined in Fig.~1 along the stripe perpendicular to the loop 
through the loop apex):
(a) highpass filtering by subtraction of a boxcar-smoothed image (top panel);
(b) baseline difference by subtraction of the first time slice (second panel);
(c) one-sided running time difference (third panel);  
(d) symmetric running time difference (fourth panel); and 
(e) minimum running time difference (bottom panel; $\Delta t = \pm 5$ frames).
There appear some ``echoes'' or ``multiple periods'', e.g., around 
$t\approx 1000$.}
\end{figure} 

\begin{figure}
\centerline{\includegraphics[width=\textwidth]{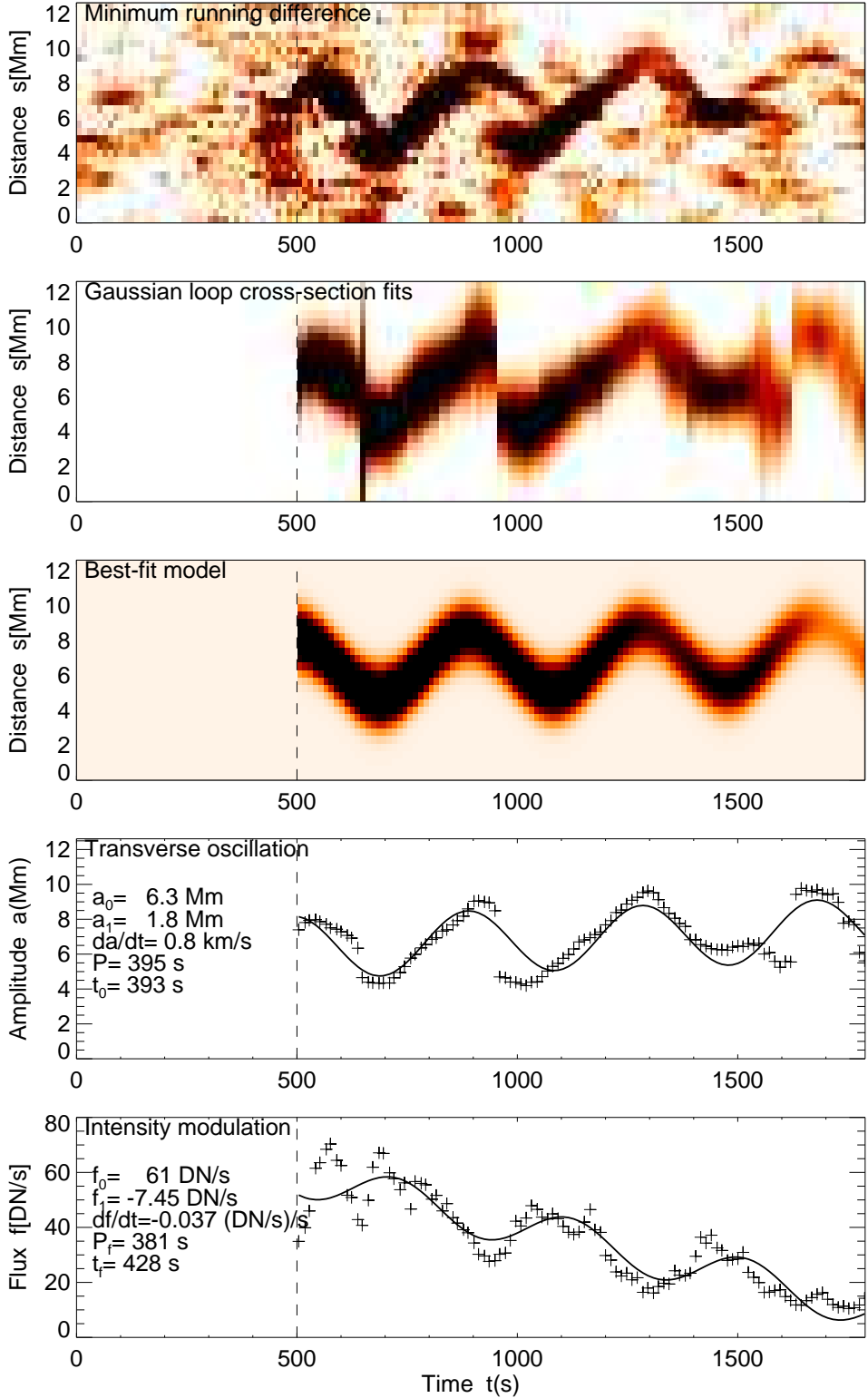}}
\caption{Time-slice diagram of 171 \ang\ flux of oscillating loop 
with the running-minimum difference method (top panel; 
$\Delta t = \pm 10$ frames), with Gaussian cross-sectional fits 
(second panel) and best-fit model (third panel), 
based on a fit of the transverse
oscillation amplitude $a(t)$ with a sine function plus linear motion 
(fourth panel), and anti-correlated flux modulation $f(t)$ at the oscillating 
loop apex (bottom panel). The data points are indicated with crosses, 
while the fit of the theoretical function is outlined with thick solid 
linestyle. The time axis is given in units of seconds after the start
of the time slice at 19:05 UT.}
\end{figure}

\begin{figure}
\centerline{\includegraphics[width=\textwidth]{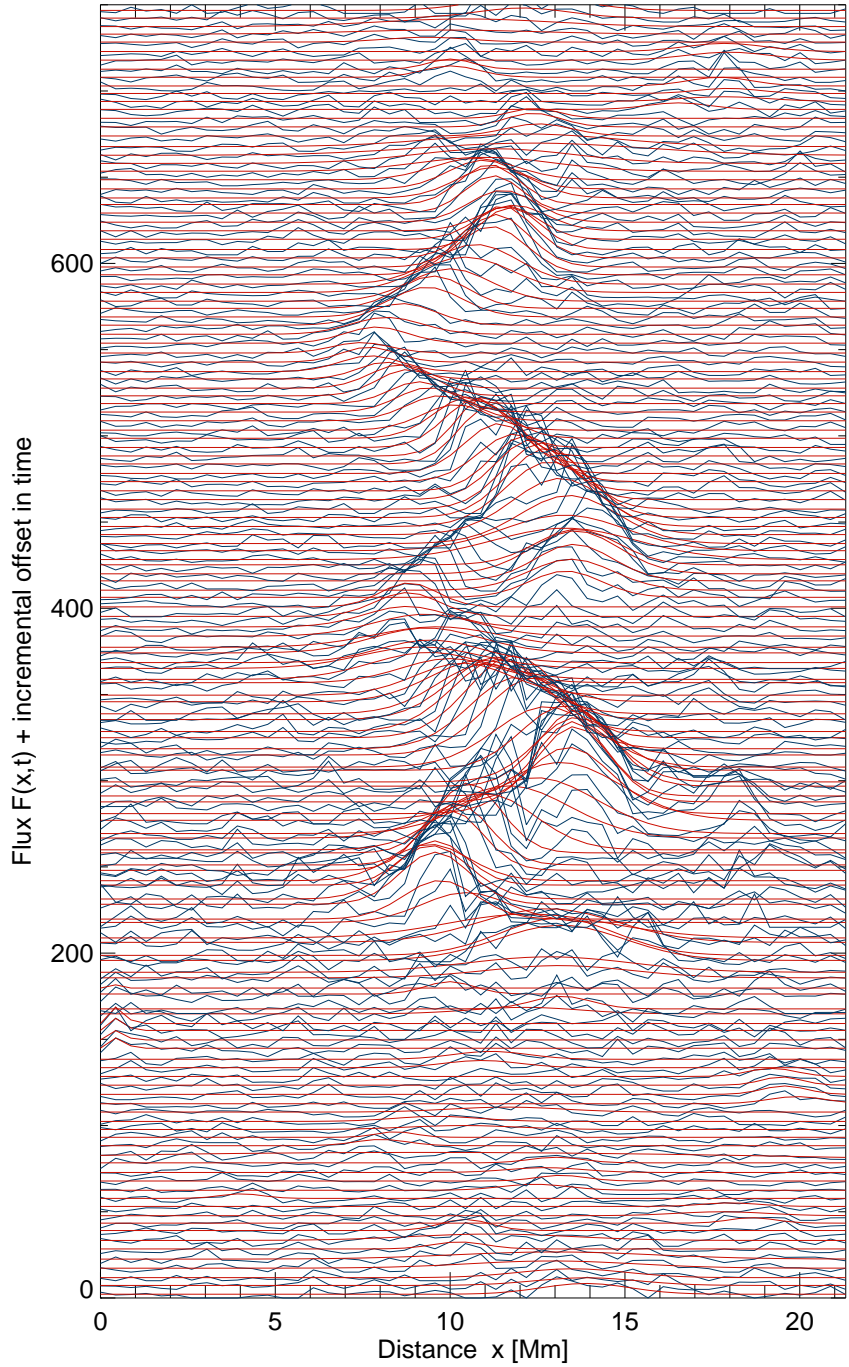}}
\caption{Cross-sectional flux profiles $F(x,t)$ (blue) obtained from 
the running-minimum difference technique (Fig.~3, top panel) and 
Gaussian fits (red).}
\end{figure}

\begin{figure}
\centerline{\includegraphics[width=\textwidth]{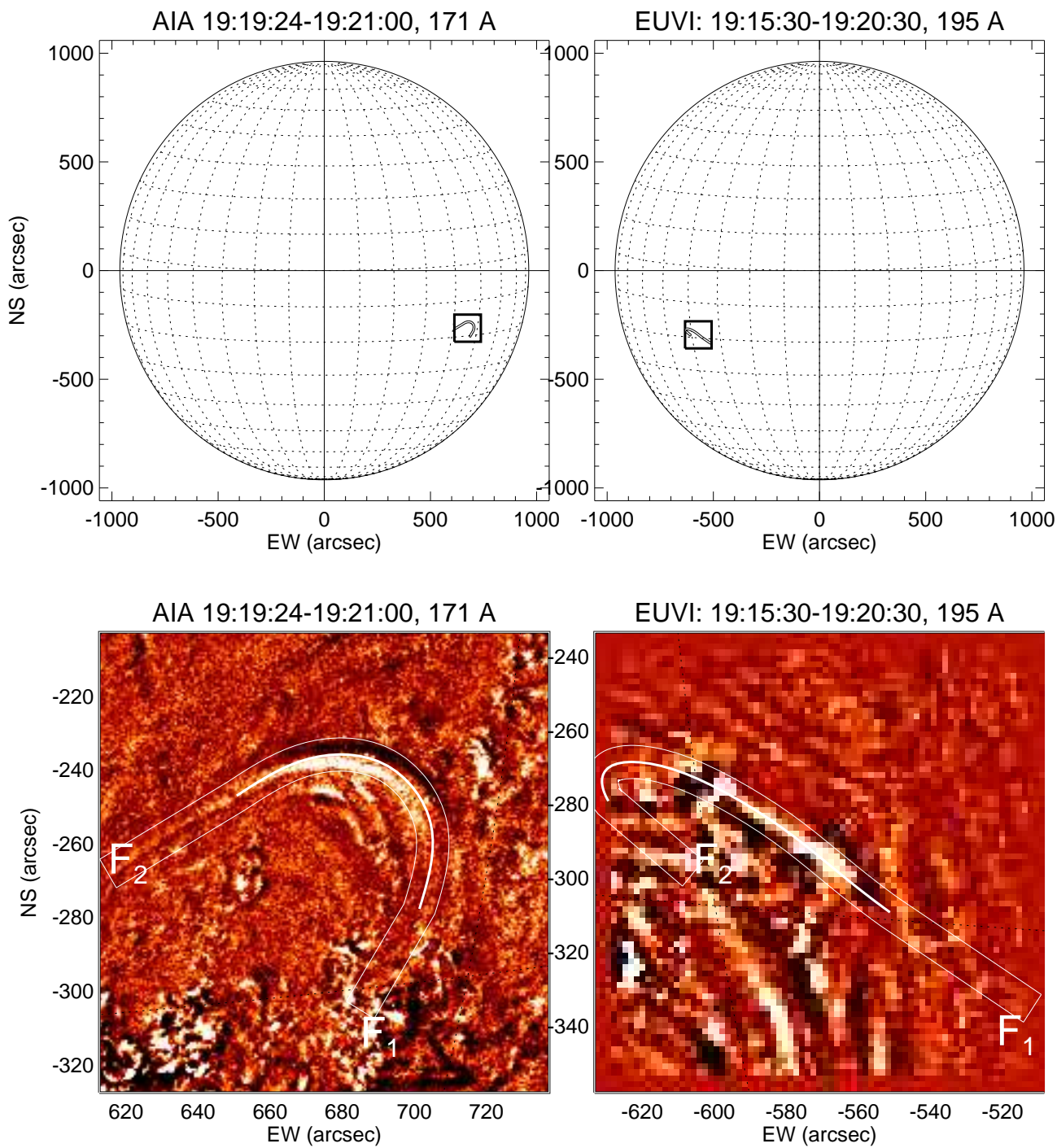}}
\caption{SDO and STEREO observations of oscillating loop:
An AIA 171 \ang\ difference image (19:19:24-19:21:00 UT) is shown
at the bottom left, and a near-simultaneous STEREO/EUVI/A 195 \ang\
difference image (19:15:30-19:20:30 UT) with additional highpass
filtering is shown at the bottom right.
The corresponding field-of-views and loop outlines are shown for
both instruments in the top panels. The thick solid curve in the
AIA image indicates the tracing of the oscillating loop segment, which
is fitted to the corresponding loop segment in EUVI/A by 3D coordinate
transformations with variable altitudes and inclination angle of the
loop plane, which constrains also the extrapolated footpoint locations 
($F_1, F_2$) at the solar surface.}
\end{figure}

\begin{figure}
\centerline{\includegraphics[width=\textwidth]{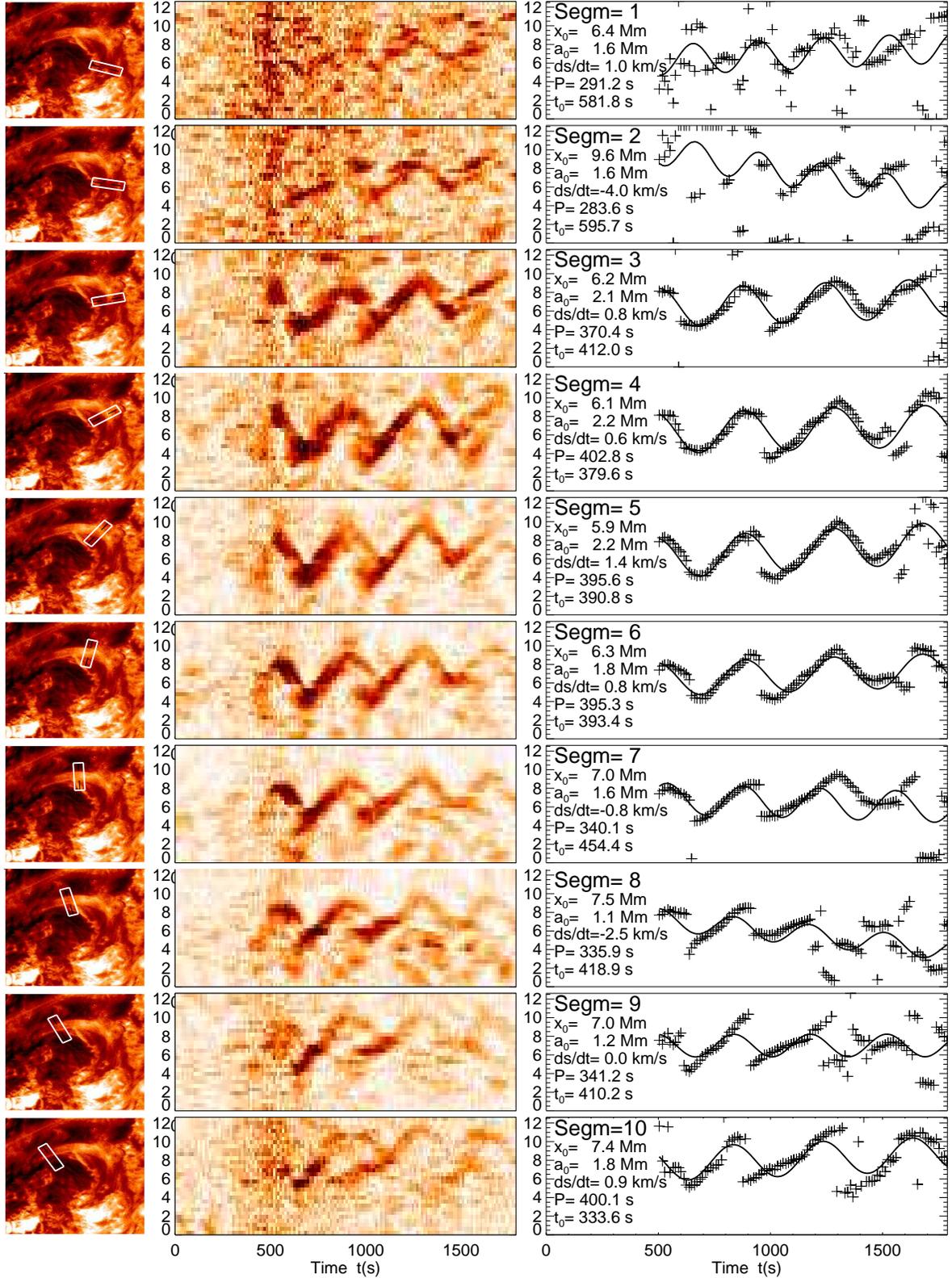}}
\caption{Ten different stripes perpendicular to the loop spine (left panels),
used to extract a running-minimum difference plot (middle panels),
and sinusoidal fits of the loop oscillation amplitude as a function
of the loop position (right panels), from loop segment 1 (near
footpoint $F_1$) to loop segment 10 (near footpoint $F_2$).}
\end{figure}

\begin{figure}
\centerline{\includegraphics[width=\textwidth]{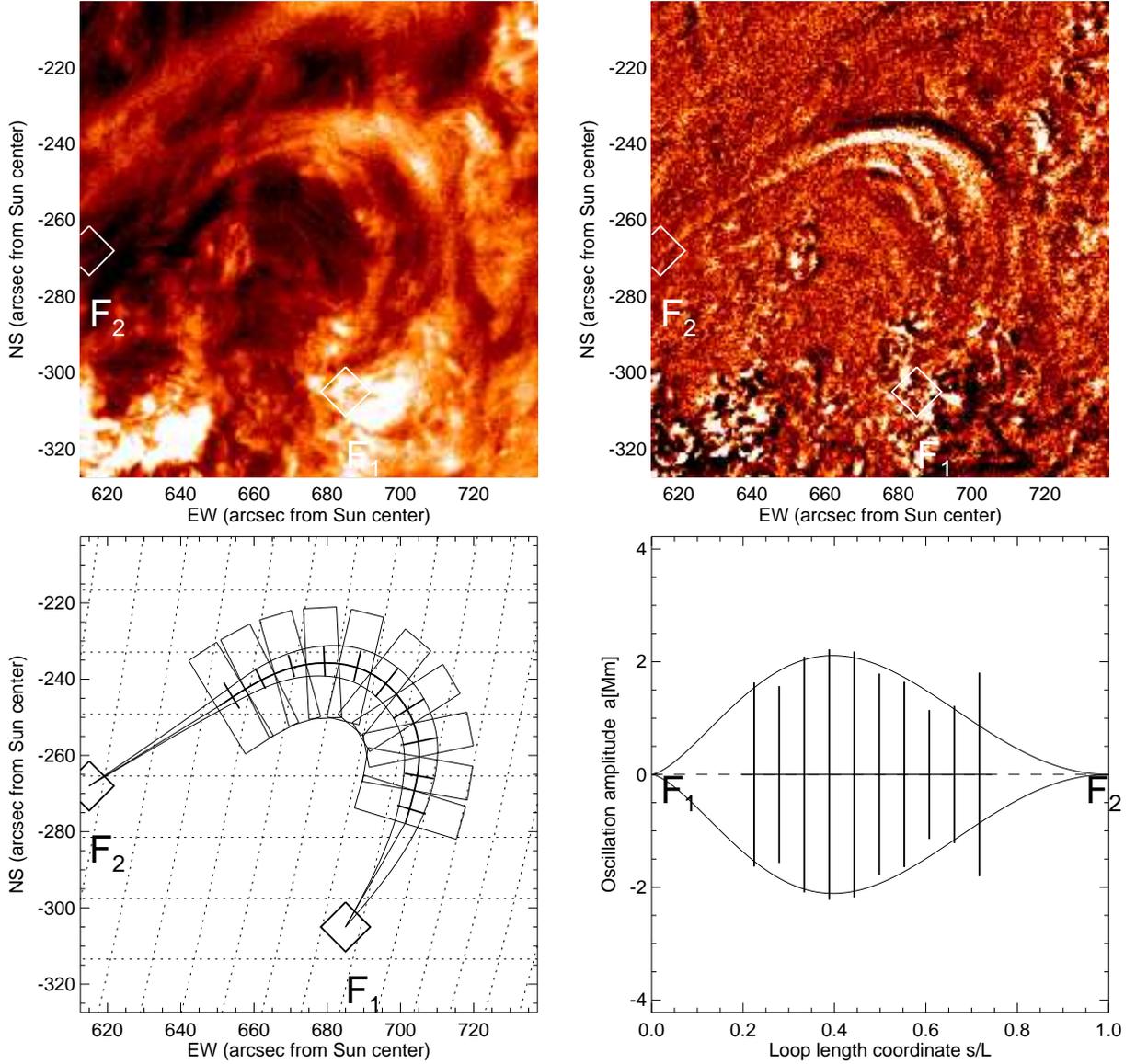}}
\caption{Location of stereoscopically triangulated loop footpoints
(marked with diamonds and labeled with $F_1$ and $F_2$) and
10 loop segments with the rectangular subimages over which the
10 time slices (shown in Fig.~6) were extracted (bottom left panel).
The corresponding AIA 171 \ang\ image (top left) and difference
image 19:21:00-19:19:24 UT (top right) are also shown.
The magnitude of the transverse kink-mode oscillation amplitude is
indicated with thick bars (bottom panels), which fit a stretched
sine function (Eq.~5).}
\end{figure}

\begin{figure}
\centerline{\includegraphics[width=\textwidth]{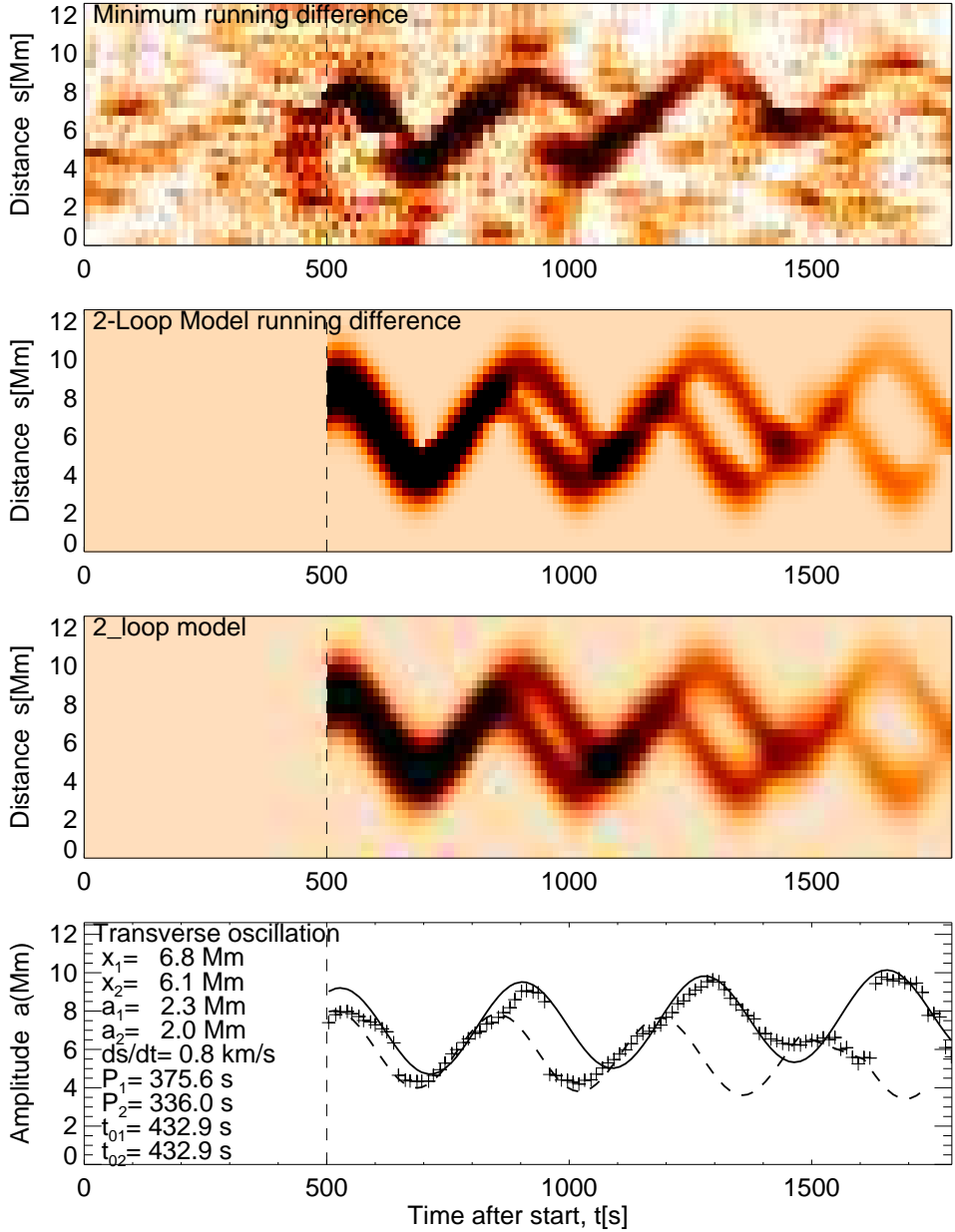}}
\caption{Two-loop oscillation model fitted to the same oscillation
amplitudes $a(t)$ as measured in Segment \#6 shown in Figs.~3 and 4.
The amplitudes of the two oscillating loops are indicated with 
solid and dashed curves (bottom panel). The two-loop solution is
also visualized as time-slice plots for the absolute flux (third
panel) and running-minimum difference representation (second panel).}
\end{figure}

\begin{figure}
\centerline{\includegraphics[width=\textwidth]{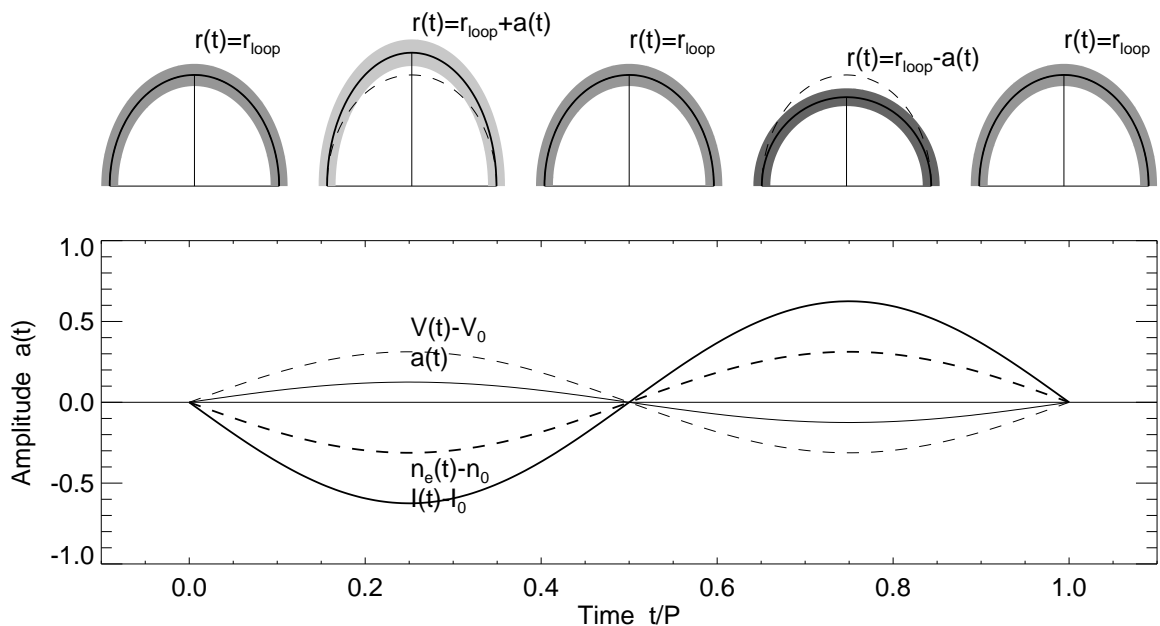}}
\caption{Schematic of sinusoidal modulation of the loop vertical radius
$r(t)=r_{loop}+a(t)$, volume $V(t)-V_0$, density $n_e(t)-n_0$, and
flux intensity $I(t)-I_0$ during one full oscillation period.
Note that the density and intensity vary in anti-correlation to the
loop amplitude, indicated with the grey-scale shading of the loops.}  
\end{figure}

\begin{figure}
\centerline{\includegraphics[width=\textwidth]{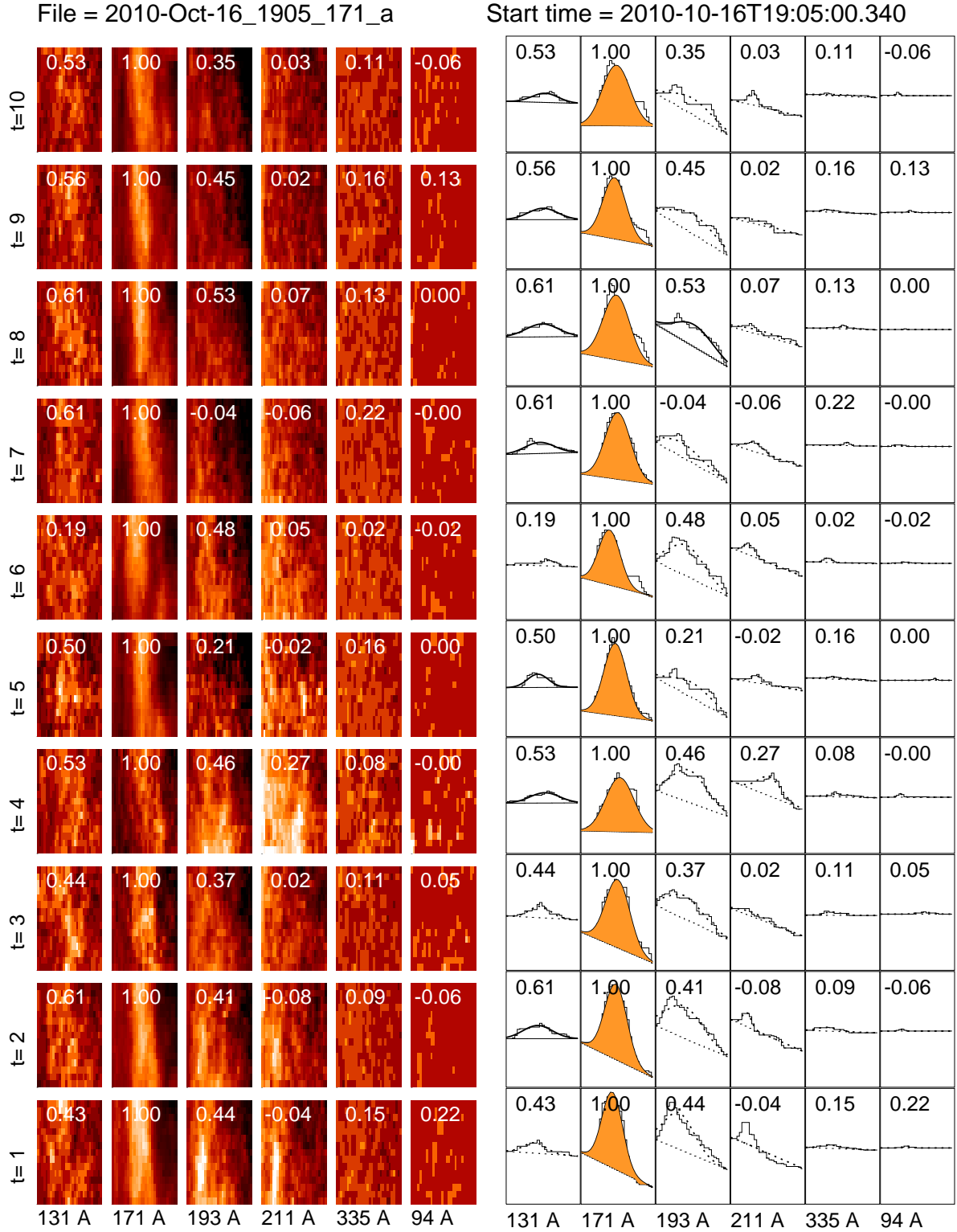}}
\caption{Cross-sectional loop profiles in 6 AIA filters and at 10 different
times during the oscillation episode from 19:05 to 19:35 UT. The wavelength
of the primary loop detection is 171 \ang , with which the subimages are
cross-scorrelated (with the cross-correlation coefficients given in each
panel).}
\end{figure}

\begin{figure}
\centerline{\includegraphics[width=\textwidth]{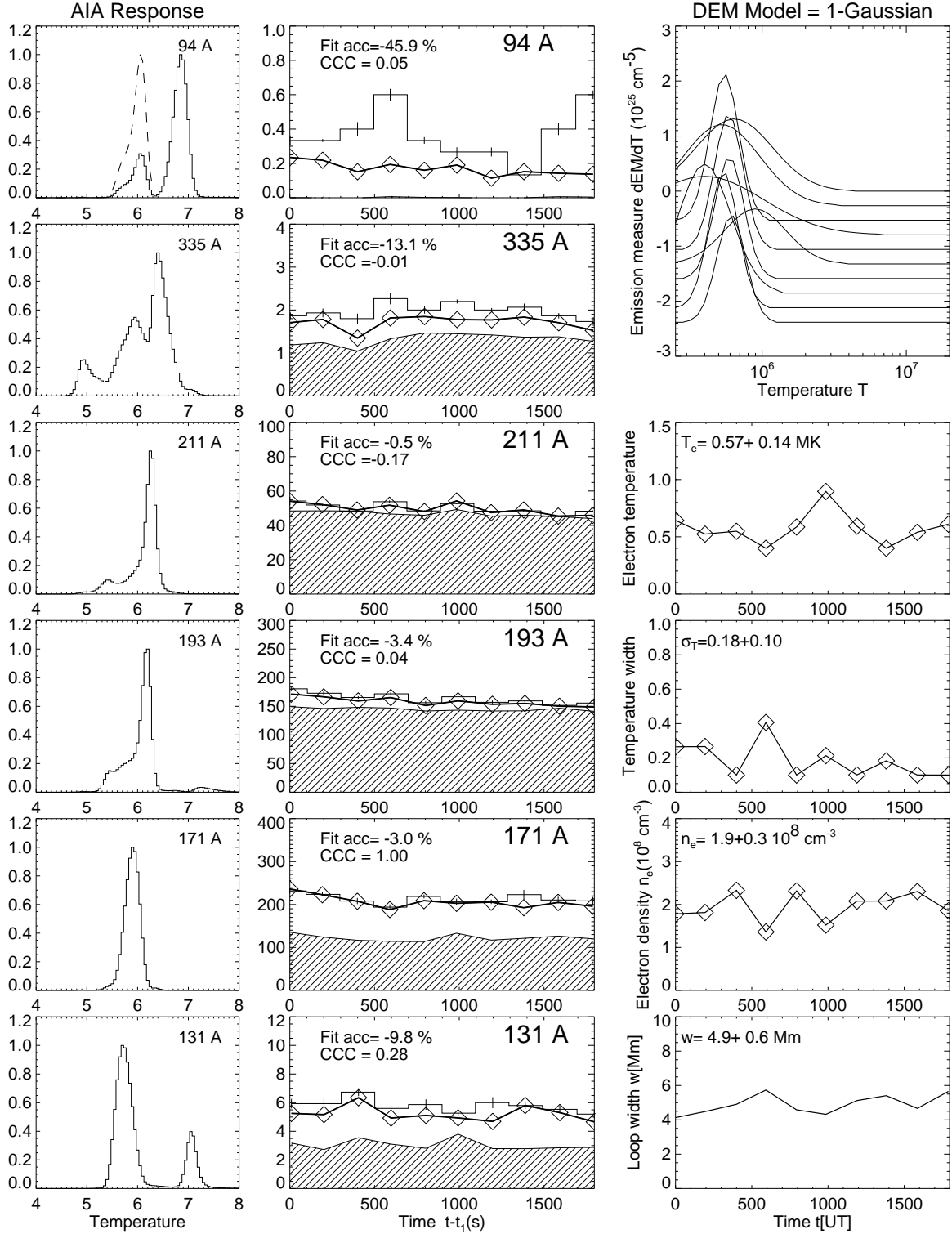}}
\caption{DEM modeling at 10 different 
times during the oscillation episode from 19:05 to 19:35 UT:
AIA response functions (left), flux versus time (middle: histogram), 
with loop background (middle hatched) and best-fit fluxes (diamonds),
DEM for 10 times (right top), temperature $T(t)$ (right second panel),
temperature width $\sigma_T(t)$ (right third panel), electron density $n_e(t)$
(right fourth panel), and loop width $w(t)$ (right bottom panel).}
\end{figure}

\begin{figure}
\centerline{\includegraphics[width=\textwidth]{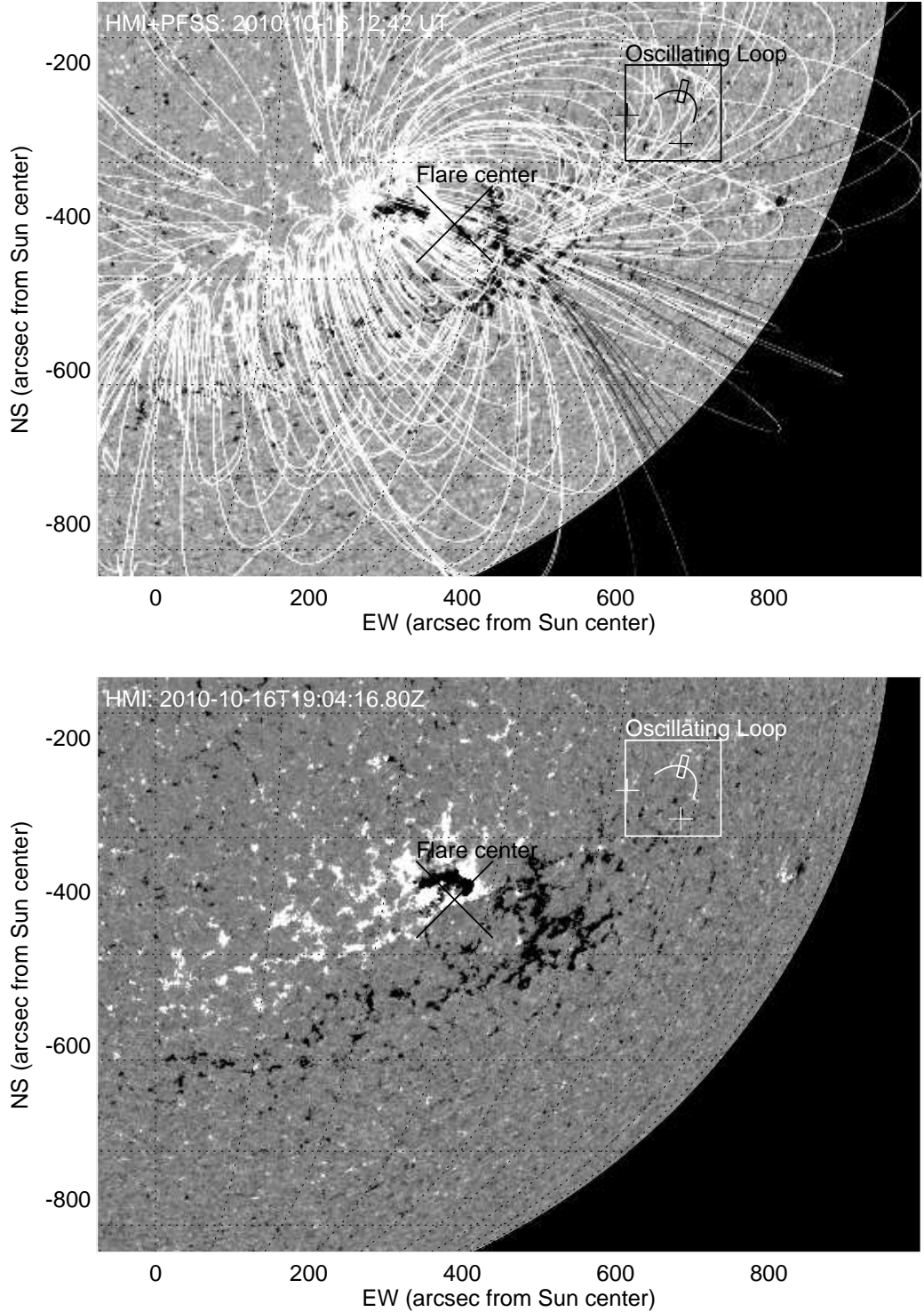}}
\caption{HMI magnetogram of same field-of-view as shown in Fig.~1,
with potential field source surface (PFSS) model field lines (top)
and locations of flare (diagonal cross), oscillating loop segment (curve),
and stereoscopically triangulated footpoints (crosses) indicated.}
\end{figure}

\begin{figure}
\centerline{\includegraphics[width=\textwidth]{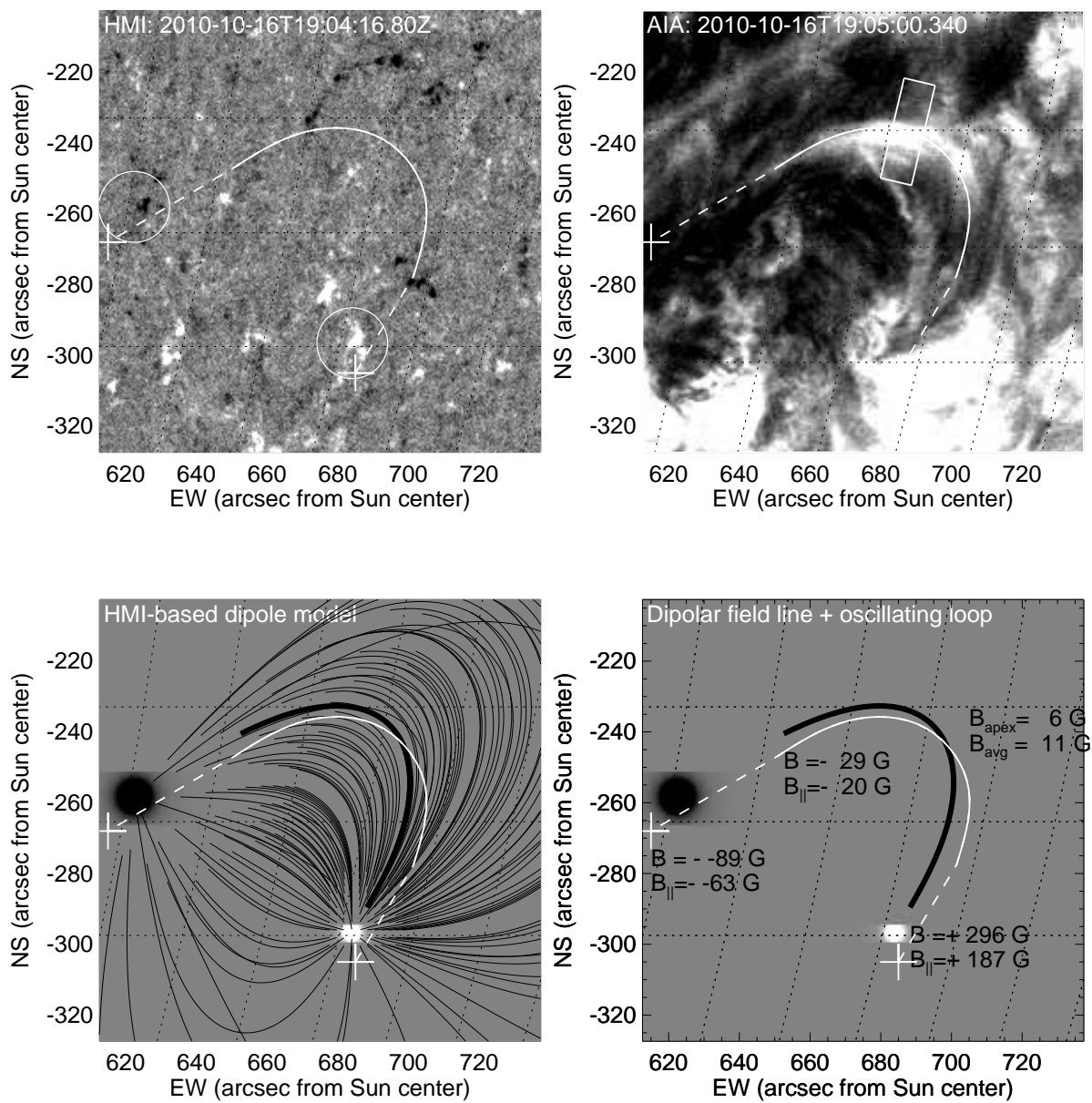}}
\caption{Enlarged field-of-view (identical to Figs.~5 and 7) of the
HMI magnetogram (top left), AIA 171 \ang\ image (top right), and
HMI-based dipolar potential field model (bottom right) of oscillating
loop (white curves). A field line that closely coincides with the 
oscillating loop is shown separately (bottom right; black curves), 
constrained by the longitudinal 
magnetic field observed in the HMI magnetograms with
$B_{\parallel 1}=187$ G and $B_{\parallel 2}=-63$ G.}
\end{figure}

\end{document}